\documentclass[11pt]{article}
\pdfoutput=1
\usepackage{jcapmod}
\usepackage{graphicx}
\usepackage{amsmath}
\usepackage{amsfonts}
\usepackage{amssymb}
\usepackage{color}
\usepackage{epsf}
\usepackage{array}
\usepackage{pifont}% http://ctan.org/pkg/pifont
\usepackage{dcolumn}
\usepackage{bm}
\usepackage{epsfig}
\usepackage{amssymb,latexsym,mathrsfs}
\usepackage{hyperref}

\newcolumntype{C}[1]{>{\centering\let\newline\\\arraybackslash\hspace{0pt}}m{#1}}

\def\beq{\begin{equation}}
\def\eeq{\end{equation}}
\def\be{\begin{equation}}
\def\ee{\end{equation}}
\def\bea{\begin{eqnarray}}
\def\eea{\end{eqnarray}}
\def\d{{\partial}}
\def\fnlequil{f_{\rm NL}^{\rm eq}}
\def\fnlloc{f_{\rm NL}^{\rm loc}}

\def\fnlorthog{f_{\rm NL}^{\rm orth}}
\def\fnlequilorthog{f_{\rm NL}^{\rm eq,\,orth}}

\def\fNL{f_{\rm NL}}

\def\mpl{M_{\rm Pl}}

\def\hinvmpc{\,h^{-1}{\rm Mpc}}

\newcommand{\ihMpc}{\ h\text{Mpc}^{-1}}

\newcommand{\gnl}{{\color{black}g_\text{NL}}}

\newcommand{\smallfoot}[1]{\let\thefootnote\relax\footnotetext{\scriptsize{#1}}}

\definecolor{darkgreen}{rgb}{0,0.5,0}
\definecolor{darkblue}{rgb}{0,0,0.75}
\definecolor{royalblue}{RGB}{0,0,128}
\definecolor{darkred}{RGB}{210,0,0}

\newcommand{\dmark}{\ding{51}\ding{51}}
\newcommand{\cmark}{\ding{51}}

\def\aap{A\&A}
\def\apj{ApJ}
\def\apjl{ApJL}

\def\mnras{MNRAS}

\def\physrep{Phys.~Rep.}
\def\prd{Phys.~Rev.~D}

\def\jcap{JCAP}

\begin{document}

\begin{titlepage}

\setcounter{page}{1} \baselineskip=14.5pt \thispagestyle{empty}

\bigskip\

%\vspace{1cm}
\begin{center}
{\fontsize{19}{36}\selectfont  \sc Testing Inflation with Large Scale Structure: \\ Connecting Hopes with Reality}
\end{center}

\begin{center}
{\fontsize{12}{15}\selectfont   \bf  Conveners: Olivier Dor\'e and Daniel Green\\
 }
\end{center}
\begin{center}
{\fontsize{12}{15}\selectfont   Marcelo Alvarez$^1$, Tobias Baldauf$^2$, J. Richard Bond$^{1,3}$, Neal Dalal$^{4}$, Roland de Putter$^{5,6}$, Olivier Dor\'e$^{5,6}$, Daniel Green$^{1,3}$, Chris Hirata$^{7}$, Zhiqi Huang$^1$, Dragan Huterer$^8$, Donghui Jeong$^9$, Matthew C. Johnson$^{10,11}$, Elisabeth Krause$^{12}$, Marilena Loverde$^{13}$, Joel Meyers$^{1}$, P. Daniel Meerburg$^{1}$, Leonardo Senatore$^{12}$, Sarah Shandera$^{9}$, Eva Silverstein$^{12}$, An\v{z}e Slosar$^{14}$, Kendrick Smith$^{11}$, Matias  Zaldarriaga$^1$, Valentin Assassi$^{15}$, Jonathan Braden$^{1}$, Amir Hajian$^{1}$,  Takeshi Kobayashi$^{1,11}$, George Stein$^1$, Alexander van Engelen$^{1}$
 }
\end{center}

%\vspace{0.2cm}

\begin{center}

\textsl{$^{1}$Canadian Institute for Theoretical Astrophysics, University of Toronto, ON}

\textsl{$^2$Institute of Advanced Studies, Princeton, NJ}

\textsl{$^{3}$Canadian Institute for Advanced Research, Toronto, ON}

\textsl{$^{4}$University of Illinois, Urbana-Champaign, IL}

\textsl{$^5$Jet Propulsion Laboratory, Pasadena, CA}

\textsl{$^6$California Institute of Technology, Pasadena, CA}

\textsl{$^{7}$Ohio State University, Columbus, OH}

\textsl{$^8$University of Michigan, Ann Arbor, MI}

\textsl{$^9$Pennsylvania State, State College, PA}

\textsl{$^{10}$York University, Toronto, ON}

\textsl{$^{11}$Perimeter Institute, Waterloo, ON}

\textsl{$^{12}$Stanford University, Stanford, CA}

\textsl{$^{13}$University of Chicago, Chicago, IL}

\textsl{$^{14}$Brookhaven National Laboratory, NY}

\textsl{$^{15}$ Cambridge University, Cambridge, UK}
\end{center} 

\hrule \vspace{0.3cm}
{ \noindent \textbf{Abstract} \\[0.2cm]
\noindent 
The statistics of primordial curvature fluctuations are our window into the period of inflation, where these fluctuations were generated. To date, the cosmic microwave background has been the dominant source of information about these perturbations. Large scale structure is however from where drastic improvements should originate. In this paper, we explain the theoretical motivations for pursuing such measurements and the challenges that lie ahead. In particular, we discuss and identify theoretical targets regarding the measurement of primordial non-Gaussianity. We argue that when quantified in terms of the local (equilateral) template amplitude $\fNL^{\rm loc}$ ($\fNL^{\rm eq}$), natural target levels of sensitivity are $\Delta \fNL^{\rm loc, eq.} \simeq 1$. We highlight that such levels are within reach of future surveys by measuring 2-, 3- and 4-point statistics of the galaxy spatial distribution. This paper summarizes a workshop held at CITA (University of Toronto) on October 23-24, 2014 \href{http://www.olivierdore.net/ng-cita.html}{[Link]}}.

 \vspace{0.3cm}
 \hrule

\vspace{0.6cm}
\end{titlepage}

\tableofcontents

\newpage

\section{Introduction}

To date, the cosmic microwave background (CMB) has been the dominant source of information about the primordial curvature perturbations.  The statistics of these fluctuations are, so far, our window into the period of inflation, where fluctuations are thought to have been generated.  However, due to the limitations set by Silk damping and foregrounds, the CMB is unlikely to offer significant improvements in the statistics of the scalar\footnote{Significant improvements in sensitivity to {\it tensor} fluctuations are expected over the next decade through measurements of CMB polarization~\cite{Abazajian:2013vfg}.  The scalar and tensor fluctuations probe different aspects of inflation and are therefore complimentary.} fluctuations (e.g. those responsible for seeding large scale structure) beyond Planck.  Large scale structure (LSS) is ultimately where drastic improvements can originate from. The purpose of this paper is to explain the theoretical motivations to pursuing such measurements and the challenges that lie ahead.

Our understanding of inflation is currently best constrained by Planck through measurements of the power spectrum~\cite{Ade:2013zuv}, bispectrum~\cite{Ade:2013ydc} and trispectrum.  The power spectrum is well characterized in terms of the amplitude, $A_s =(2.23 \pm 0.16) \times 10^{-9}$ and the tilt $n_s =0.9616\pm0.0094$.  These values can give us insight into the underlying inflationary background.  The bispectrum can be used to test a wide variety of models beyond the single-field slow-roll paradigm.  Often these results are reported in terms of the local template~\cite{Salopek:1990jq, Komatsu:2003iq}, $\fNL^{\rm loc} =2.7 \pm 5.8$, the equilateral template~\cite{Creminelli:2005hu}, $\fnlequil = -42\pm 75$, and the ortogonal template~\cite{Senatore:2009gt}, $\fNL^{\rm orthogonal} = -25 \pm 39$ .  Similar constraints have been placed on several trispectrum templates \cite{Fergusson:2010gn, Sekiguchi:2013hza}.  While these results are impressive, we should ask if they have reached a theoretically desirable level of sensitivity.  In other words, one should put these results into the context of what natural levels of non-Gaussianity might be expected under plausible theoretical expectations.

In the specific context of the bispectrum, the local and equilateral templates probe qualitatively different deformations of  single-field slow-roll inflation and should be assessed independently.  Local non-Gaussianity is a sensitive probe of multi-field models due to the single-field consistency relations~\cite{Maldacena:2002vr,Creminelli:2004yq}.  Equilateral non-Gaussianity is a probe of non-slow roll dynamics through self-interactions of the perturbations.  Despite the clear differences between these probes, we will argue that a natural target level of sensitivity is $\Delta \fNL^{\rm loc, eq.} < 1$.  Given the target level of sensitivity, there is still significant room for improvement beyond the limits set by the CMB.  This presents a well-defined challenge for upcoming LSS surveys.

The state of the art for testing non-Gaussianity in LSS is constraining the scale-dependent halo bias predicted for models with $\fnlloc$~\cite{Dalal:2007cu}.  These constraints can be derived from the halo power spectrum which greatly simplifies the analysis.  However, these measurements are prone to systematic effects which present a significant barrier for existing surveys to making this measurement competitive with the CMB.  In addition, the halo bispectrum would be a necessary tool for testing non-local type non-Gaussianity like $\fnlequil$. To date, there is no constraint on primordial non-Gaussianity from the halo bispectrum, despite suggestions that it contains substantially more signal to noise than the halo power spectrum~\cite{Baldauf:2010vn}.  Ultimately, the bispectrum will need to be used to make LSS as versatile a tool as the CMB has become.

The goal of this paper is to summarize the status, motivations and challenges for testing inflation in large scale structure surveys.  Our discussion will be somewhat weighted towards $\fnlloc$ because it has been demonstrated that large scale structure can give comparable constraints to the CMB.  However, we will emphasize that there are a number of other interesting theoretical targets that deserve equal attention in future surveys that will ultimately be tested through probes other than scale dependent bias. This paper aims at sharing with the community the outcome of our workshop and thus should not be read as a review. Many relevant subjects were not discussed for lack of time and many relevant references are missing. 

The paper is organized as follows.  In section \ref{sec:theory}, we will review a number of theoretical targets that would be desirable to meet in future surveys.  In particular, we will explain why $\Delta \fNL^{\rm loc, eq.} < 1$ are particularly interesting levels of sensitivity.  In section \ref{sec:local}, we will discuss the status and prospects for measuring $\fnlloc \sim 1$ in future surveys.  In section \ref{sec:bispec}, we will explain the status of the bispectrum as a tool for testing inflation.  In section \ref{sec:sys}, we will discuss systematics that may be relevant to making these measurements a reality.  

%Olivier: please be sure somebody also mentions the power spectrum measurement. You can look at pag. 14 of 1309.4060 . I think this importance of the low-k power spectrum measurement in LSS for the change of slope of the spotential has been rarely pointed out, and it is pretty important.

\section{Theoretical motivation and targets for non-Gaussianity}\label{sec:theory}

The size and form of primordial non-Gaussianities in single-clock inflationary models (i.e.~models where only one light dynamical field is relevant during inflation) has several remarkable properties intimately related to their nature. These properties can be seen by studying the effect of large modes on smaller scales ones as inflation proceeds.  This corresponds to the so-called squeezed limit of correlation functions. Consider how a mode of momentum $q$ affects a mode of momentum $k$ with $q\ll k$. The mode $q$ leaves the horizon much before the mode $k$ and becomes part of the background in which the mode $k$ will freeze. 

One of the basic properties of inflation is that the inflationary background is an attractor. As a result, the effects of the mode $q$ quickly become unobservable and the mode $k$ freezes in a background that is basically indistinguishable from the unperturbed background. In fact the effects of the $q$ mode redshifts as $(q/aH)^2$. Thus one expects any effect of the long mode on the short to be suppressed by $(q/k)^2$. 

This physical  fact  is encoded in the various consistency relations that $N$-point functions satisfy in single clock inflation~\cite{Maldacena:2002vr,Creminelli:2004yq}.  For example if one considered the three point function with momenta $q$, $k_1$ and $k_2$ such that $q \ll k_1\sim k_2 = k$ the leading piece of the three point function is: 
\begin{equation}
\langle\zeta(q) \zeta(k_1)\zeta(k_2)\rangle^\prime \sim \Big[ (n_s-1)  + {\cal O}\Big(\frac{q^2}{k^2}\Big)\Big] P(q) P(k) \ ,
\end{equation}
where the prime indicates that we are suppressing the momentum conserving delta-function. This piece, proportional to $(n_s-1)$, results form the fact that the long mode is not affecting the short scales and is non-zero due to our choice of coordinates. In the background of a long mode the comoving scale $k$ corresponds  to physical a scale $k_F$ given by $k = e^{\zeta_L} k_F$.  The  $(n_s-1)$ factor in the three point function is there to guarantee that the amplitude of fluctuations on given physical scales are independent of the long mode. Schematically:
\begin{equation}
\langle\zeta(q) \zeta(k_{F1})\zeta(k_{F2})\rangle^\prime = {\cal O}\Big(\frac{q^2}{k^2}\Big)\ .
\end{equation}
The vanishing of the effect of long modes on short ones is a consequence of the attractor nature of the inflationary solution and of the fact that modes freeze and become part of the background sequentially. The mode $q$ is ``hidden" from modes that freeze later\footnote{This expectation can be modified if the attractor phase of single clock inflation lasts for the minimal number of e-folds and is preceded by a non-attractor phase~\cite{Namjoo:2012aa,Martin:2012pe,Chen:2013aj} or by some physics captured by a non-Bunch Davies initial state for the fluctuations~\cite{Chen:2006nt, Holman:2007na, Agullo:2010ws,Ganc:2011dy}.  Nevertheless, these scenarios are often observationally distinguishable from single or multi-field inflation~\cite{Flauger:2013hra}.}.

The second remarkable property of the non-Gaussianities in single-field {\it slow-roll} models is how small even the $ O(q^2/k^2)$ piece is~\cite{Maldacena:2002vr}. In fact, the full result scales as 
\begin{equation}
\langle\zeta(q) \zeta(k_1)\zeta(k_2)\rangle^\prime \sim \Big[ (n_s-1)  +\epsilon {\cal O}\Big(\frac{q^2}{k^2}\Big)\Big] P(q) P(k) \ ,
\end{equation}
with $\epsilon=-\dot H/H^2$ during inflation. The fact that the $q^2/k^2$ piece is down by a factor of $\epsilon$  is again directly related to the mechanism that single clock inflation uses to create the fluctuations, the fact that they are time-delay fluctuations. 

To highlight this fact, consider the effect of long modes on short ones in the late universe, say during the matter era. Long modes can be thought of as producing a curved universe with a curvature parameter $\Omega_K \sim q^2 \zeta / a^2 H^2$ on which the short modes evolve~\cite{BaldaufEtal,Creminelli:2013cga}. This curvature changes the growth factor of perturbations. In the presence of the long modes the growth rate is changed and the coefficient relating that change to $\Omega_K$ is of order one. However during {\it slow-roll} inflation, the effect of the curvature of the long mode is not of order one, but of order $\epsilon$. This reflects the fact that in slow-roll inflation one is dealing just with time delay fluctuations, and that even if the surfaces of constant value of the clock are curved, this is not changing the space-time. More precisely those changes are down by $\epsilon$. 

Models in which the perturbations are originally not perturbations of the inflationary clock, but are perturbations of another field that are then converted into curvature fluctuations, behave very differently~\cite{Byrnes:2010em,Suyama:2013nva}. The conversion requires that super-horizon modes affect the equation of state of matter so as to make different regions of the Universe expand by different amounts. Thus, super-horizon modes in these scenarios almost by definition must produce locally observable effects. Non-linearities in the Einstein equations and the relation between field fluctuations and the changes in the equation of state will couple all modes during the conversion process. As a result, in these models there will be violations of the consistency condition. It is important to stress that these considerations are quite general and do not depend on whether fluctuations are generated during a period of inflation or about some other background. 

We will review some illustrative examples in the next section. The general situation is that various non-linearities in the conversion process result in order one coupling between modes. Thus if the conversion to curvature fluctuations is efficient then one expects $\fNL^{\rm loc} \sim 1$. If, however, the conversion is somewhat inefficient, say if $10^{-4}$ fluctuations in the second field are needed to create $10^{-5}$ curvature fluctuations, then one expects $\fNL^{\rm loc} \sim 10$. The Planck constraints have already shown that the conversion mechanism has to be efficient. It is thus clear that $\fNL^{\rm loc} \sim 1$ is a clear target as even assuming efficient conversion one gets contributions of this level. The target is not sharp because there is more than one contribution that can be balanced to produce a partial cancellation.  In addition, the curvature perturbation from the conversion may be sub-dominant to the perturbations of the clock, in which case $\fnlloc< 1$ is also plausible.

Thus we conclude that the structure of the correlations between modes in the squeezed limit --the fact that the correlations are so small-- is intimately linked with the basic properties of single-field models. The sequential hiding of fluctuations as part of the background which is an attractor makes the local contribution to $N$-point functions vanish. The fact that fluctuations are just time-delay fluctuations further suppresses the piece of order $q^2$. Thus explicitly checking that both of these contributions are significantly smaller than unity would provide nice evidence in favor of this picture.

\subsection{Local Non-Gaussianity ($\fnlloc$)}

To better illustrate what level of precision of observational constraints are theoretically interesting, it is useful to consider some specific models.  Our focus will be on the simplest examples of a class of models known as `spectator field' models, where inflation is driven by a field $\phi$, and there exists another field $\sigma$ whose energy density is subdominant during inflation and whose fluctuations are primarily responsible for the curvature fluctuations which are observed in the cosmic microwave background and large scale structure.  This class of models certainly does not cover the full set of possibilities for multiple field inflation, though it provides a useful set of examples whose predictions can be contrasted with those of single field inflation.  For reviews of multiple field inflation models, including several which are not discussed here, see for example \cite{Byrnes:2010em,Suyama:2013nva}.

\subsubsection{Curvaton Scenario}
In the curvaton scenario \cite{Linde:1996gt,Enqvist:2001zp,Lyth:2001nq,Moroi:2001ct} the spectator field $\sigma$ is light and nearly frozen in its potential until after the end of inflation.  After inflation, the field $\phi$ which drove inflation decays into radiation which redshifts as $a^{-4}$.  When the Hubble rate drops below some specific value, the field $\sigma$ begins oscillating about the minimum of its potential which, if the potential is quadratic, causes the energy density of the field $\sigma$ to decrease as $a^{-3}$.  As a result, while the energy density of $\sigma$ was negligible during inflation, it comes to make up a significant fraction of the energy density after inflation.  At some point the field $\sigma$ decays into radiation which thermalizes with the decay products of $\phi$, and the final radiation fluid acquires fluctuations which depend upon the initial fluctuations of the field $\sigma$.

There are two key parameters which determine the value of local non-Gaussianities in the curvaton scenario.  The first is the ratio of the energy density in the field $\sigma$ compared to radiation at the time of the decay of $\sigma$, conveniently defined through the parameter $r_\mathrm{dec}$ given by
\begin{equation}
	r_\mathrm{dec}\equiv\left.\frac{3\rho_\sigma}{3\rho_\sigma+4\rho_r}\right|_{t=t_\mathrm{dec}} \, ,
\end{equation}
which is restricted to be in the range $0\leq r_\mathrm{dec} \leq 1$. We will also define a second parameter, $g(\sigma_*)$, that relates the energy density of $\sigma$ at the onset of the oscillating phase to its initial field value:
\begin{equation}
	\rho_\sigma=\frac{1}{2}m_\sigma^2g(\sigma_*)^2 \, .
\end{equation}
In terms of these parameters, the curvaton scenario predicts \cite{Sasaki:2006kq}
\begin{equation}
	f_{\mathrm{NL}}^\mathrm{loc}=\left(\frac{\mathcal{P}_\zeta^\sigma}{\mathcal{P}_\zeta^\sigma+\mathcal{P}_\zeta^\phi}\right)^2\left[\frac{5}{4r_\mathrm{dec}}\left(1-\frac{gg''}{{g'}^2}\right)-\frac{5}{3}-\frac{5r_\mathrm{dec}}{6}\right] \, ,
\end{equation}
where $\mathcal{P}_\zeta^\sigma$ is the contribution to the power spectrum from $\sigma$, and likewise for $\phi$.  We see that in the spectator regime, where $\mathcal{P}_\zeta^\phi\ll \mathcal{P}_\zeta^\sigma$, the curvaton scenario typically predicts $\left|f_{\mathrm{NL}}^\mathrm{loc}\right|\geq\mathcal{O}(1)$.

\subsubsection{Modulated Reheating}\label{sec:modreh}
The energy density of the field $\phi$ which drove inflation must be converted into radiation after inflation ends.  In the modulated reheating scenario \cite{Dvali:2003em,Kofman:2003nx,Dvali:2003ar}, the rate at which this transfer of energy occurs is controlled by the value of the spectator field $\sigma$.  The spatial variations of the field $\sigma$ give rise to spatial variations in the expansion history, thus generating curvature perturbations.  In the limit that the decay rate is small compared to the Hubble rate at the end of inflation, the modulated reheating scenario predicts
\begin{equation}
	f_{\mathrm{NL}}^\mathrm{loc}=\left(\frac{\mathcal{P}_\zeta^\sigma}{\mathcal{P}_\zeta^\sigma+\mathcal{P}_\zeta^\phi}\right)^2\left[5\left(1-\frac{\Gamma\Gamma''}{{\Gamma'}^2}\right)\right] \, ,
\end{equation}
where $\Gamma$ is the decay rate of the inflaton and $\Gamma'\equiv\partial \Gamma(\sigma)/\partial \sigma_*$ gives the dependence on the spectator field. We again find that in the spectator regime $\left|f_{\mathrm{NL}}^\mathrm{loc}\right|\geq\mathcal{O}(1)$.

\subsubsection{Discussion}
While we discussed only two examples here, the statement that spectator models tend to predict $\left|f_{\mathrm{NL}}^\mathrm{loc}\right|\geq\mathcal{O}(1)$ applies in several other cases as well \cite{Suyama:2013nva}.  This leads us to the rough conclusion that an observational constraint at the level of $\Delta f_{\mathrm{NL}}^\mathrm{loc}\simeq\mathcal{O}(1)$ is of particular theoretical interest.  Specifically, observations which reveal that $\left|f_{\mathrm{NL}}^\mathrm{loc}\right|\leq\mathcal{O}(1)$ would tend to disfavor spectator models apart from those with special choices of model parameters.  Put another way, such a constraint on $f_{\mathrm{NL}}^\mathrm{loc}$ would favor a model of the early universe where the fluctuations in the field whose energy density drove inflation cannot be neglected.  This of course would not rule out multiple field inflation in general, since observable non-Gaussianity is not a general prediction of all such models \cite{Vernizzi:2006ve,Byrnes:2008wi,Meyers:2010rg,Elliston:2011dr,Meyers:2013gua}, though improved observational bounds also help to constrain these more general scenarios.

Alternatives to inflation would be more strongly constrained by such a measurement because the fluctuations of spectator fields are necessary to produce a scale-invariant spectrum~\cite{Baumann:2011dt}.  As a result, constrains of $\left|f_{\mathrm{NL}}^\mathrm{loc}\right|\leq\mathcal{O}(1)$ would rule out most of the space of viable alternatives to inflation (see e.g.~\cite{Ijjas:2014fja}).

Finally, models with non-Gaussianity that couples modes of very different wavelengths, as the $f_{\rm NL}^{\rm loc}$ does, have the feature that amplitude of the fluctuations and of the non-Gaussianity can be substantially different in different spatial subvolumes whose long-wavelength background modes do not take the mean value (zero). Within our universe this is a useful feature for constraining non-Gaussianity, either through halo bias (discussed in Section \ref{sec:local}) or through a more general position dependent power spectrum \cite{Chiang:2014oga}. However, this effect also means that for models that predict more than the minimal number of e-folds there is a new source of cosmic variance in using any observed amplitude of non-Gaussianity to constrain the parameters of the model \cite{Nelson:2012sb,Nurmi:2013xv,LoVerde:2013xka}. If $f_{\rm NL}^{\rm loc}$ can be observationally constrained to be less than 1 on a sufficient range of scales, this additional cosmic variance is likely to be irrelevant for comparing observations to theory \cite{Nelson:2012sb, Bramante:2013moa}.

\subsection{Equilateral and Othorgonal Non-Gaussanity ($\fnlequil, \fnlorthog$)}

Single field inflation can produce non-Gaussian shapes of the form of $\fnlequil$ and $\fnlorthog$.  They are common signatures of models with dynamics beyond the single-field slow-roll inflation (see e.g.~\cite{ArmendarizPicon:1999rj,Alishahiha:2004eh, ArkaniHamed:2003uz, Chen:2006nt, Green:2009ds, Barnaby:2010vf, LopezNacir:2011kk}).  As in the case of $\fnlloc$, we would like to know if there is a natural threshold value for these parameters such that a non-detection at such a level would represent a major improvement in our knowledge of inflation? We now explain that such a threshold is indeed $\fnlequil\sim 1$ and $\fnlorthog\sim 1$. 

In a modern way of thinking of inflation, single field inflation is considered in much more general terms than standard slow-roll inflation.  Instead, inflation can be studied as the theory of the fluctuations of spontaneously broken time translations around a quasi de Sitter background. This is the so-called Effective Field Theory of inflation~\cite{Cheung:2007st}; its action, after some useful rescalings of the field and the coordinates, reads
\be
S=\int d^4x\;\sqrt{-g}\;\left[\left(\d_\mu\pi_c\right)^2+\frac{\dot\pi_c(\d_i\pi_c)^2}{\Lambda_1^2}+\frac{\dot\pi_c^3}{\Lambda_2^2}+\ldots\right] \ .
\ee
The field $\pi_c$ represents the canonically-normalized  Goldstone boson of time-translations, and the curvature perturbation $\zeta$ that is constant of super-Hubble scale is related to $\pi_c$ on super-Hubble scales as $\zeta=-[H/(2\dot H\mpl^2 c_s)^{1/2}]\pi_c$, where $c_s$ is the speed of sound of the fluctuations.

The scales $\Lambda_{1,2}^4$ are of order $\Lambda^4\sim (\dot H\mpl^2 c_s^5)$. They are related to the perturbative unitarity bound of the theory: the theory becomes strongly coupled larger energy scales, and the theory is therefore modified above those thresholds~(see e.g.~\cite{Cheung:2007st, Baumann:2011su,Baumann:2014cja} for further details). The parameters $\fnlequil$ and $\fnlorthog$ can be expressed in terms of $\Lambda$ as
\be
\fnlequil\zeta\sim\fnlorthog\zeta\sim \frac{H^2}{\Lambda^2}\ .
\ee
Improving limits on the teo parameters $f_{\rm NL}$ parameters is therefore equivalent to increasing the hierarchy between $\Lambda$ and $H$. Note that something very interesting happens if we push $\fnlequilorthog\lesssim 1$; in that case, $\Lambda^4\gtrsim \dot H\mpl^2$, corresponding to the same energy scale as the kinetic energy of the scalar field in standard slow-roll inflation. In fact, if we consider slow-roll inflationary models with self interactions
\be
S=\int d^4x\;\sqrt{-g}\;\left[\frac{1}{2}(\d_\mu\phi)^2+V(\phi)+\frac{(\d_\mu\phi)^4}{\Lambda_\phi^4}\right]\ ,
\ee
we can ask what is the maximum value of $\fnlequilorthog$ compatible with the fact that we wish to have a slow rolling solution where the higher derivative term is not important for the background solution $\dot\phi_{\rm sr}\sim V'/H$~\cite{Creminelli:2003iq}. The requirement that the background solution is unaffected by the higher derivative term implies    $\Lambda_\phi^4\gtrsim \dot\phi^2_{\rm sr}$. In turns, this implies that the produced $\fnlequilorthog$ by the corresponding interaction is less than one.

All of this tells us that if we can confirm that the values of $\fnlequil$ and $\fnlorthog$ to be less than one, then the inflationary theory is so weakly interacting that it can be described as a small perturbation of slow roll inflation, which we can describe in great detail. In other words, while current observations allow inflation to have wildly different dynamics, by constraining $\fnlequil$ and $\fnlorthog$ to less than one we will have cornered inflation to be of the slow-roll kind\footnote{As for most of theorems, there are some caveats. First, it could be that the leading interactions in single-field inflation induce a four-point function, and not a three-point function~\cite{Senatore:2010jy}.  In this case, the lower bound on $\Lambda$ from the same data set would be lower. Additionally, it could be that the leading interactions that produce a three-point function are higher-derivatives, a fact that makes the induced limit on $\Lambda$ smaller~\cite{Creminelli:2010qf,Behbahani:2014upa}. Finally, the measurement of the scale $\Lambda$ is indirect, coming from measurements at energy scales of order $H$, not of order $\Lambda$. This means that some phase transition might actually occur when we connect the theory around the inflationary background, and that is what we measure, to the slow-roll inflation model, which is defined around the Minkowski vacuum. Such a possibility is difficult to exclude from observations at energies $H\ll\Lambda$. Even though these would remain allowed possibilities even if $\fnlequilorthog$ are pushed below one, we regard them as rather exotic. }. Together with constraining $\fnlloc\lesssim 1$, alternatives to single-field slow-roll inflation would be disfavored. This would represent a major step forward, that could be achieved even without a positive detection, by simply setting an upper limit on some observables. In contrast, a positive detection would tell us that we are far away from the single-field slow-roll regime, a fact that opens a plethora of interesting theoretical and observational possibilities. 

\subsection{Intermediate Shapes}

In many models of inflation there are additional fields present that decay outside the horizon.  As a result, they are diluted before the end of inflation and do not modify the reheating surface.  In these models, the curvature perturbation is still the fluctuations of the inflationary clock but the freeze-out of the clock can be modified by the presence of these additional fields.  A canonical example of the type is quasi-single field inflation~\cite{Chen:2009zp, Chen:2009we} and generalizations thereof~\cite{Baumann:2011nk, Noumi:2012vr, Green:2013rd}.  Most significantly, these additional fields can modify the behavior of the bispectrum in the squeezed limit, violating the single field consistency conditions.  The resulting behavior lies somewhere between local and equilateral shapes without large violations of scale invariance.  For example, when the inflaton is coupled to a scalar of mass $m$ satisfying $\frac{3}{2} H > m > 0$, a squeezed bispectrum is generated of the form
\begin{equation}
\langle\zeta(q) \zeta(k_1)\zeta(k_2)\rangle_{q\ll k_1\sim k_2}^\prime \sim \Big[ (n_s-1)   + {\cal O}\Big(\frac{q^\alpha}{k^\alpha}\Big) \Big] P(q) P(k)  \ ,
\end{equation}
where $\alpha = 3/2 - \sqrt{9/4 - m^2/H^2}$.  These models produce a squeezed signature that is distinguishable from single-field models but is not of the local type.

One particularly interesting feature of these scenarios is that one can produce large non-Gaussianity with relatively weak couplings between the inflaton and these extra degrees of freedom~\cite{Green:2013rd, Assassi:2013gxa}.  This makes these intermediate shapes a compelling target from a particle physics perspective as large numbers of additional fields are very plausible at the energy scales relevant to inflation~\cite{Baumann:2014nda,Craig:2014rta}.  

\subsection{Theory targets summary}

We can summarize our findings in the following table:\vspace{0.5cm}
\begin{table}[!h]
\begin{center}
\begin{tabular}{||c|c|c||}
\hline \hline
\rule[-2mm]{0mm}{6mm} & $\fnlloc\lesssim 1$ & $ \fnlloc\gtrsim 1$ \\
\hline\hline
\rule[-2mm]{0mm}{6mm} $\fnlequilorthog\lesssim 1$ & Single-field slow-roll  & Multi-field\\
\hline
\rule[-2mm]{0mm}{6mm} $\fnlequilorthog\gtrsim 1 $ & Single-field non-slow-roll & Multi-field \\
\hline\hline
\end{tabular}
\end{center}
\caption{Table summarizing physical implications for qualitatively different measurements of
  the shapes of primordial non-Gaussianity.}
\label{tab:summary}
\end{table}

As emphasized above, the interpretation of each scenario requires some caveats.  It is our assessment that this table represents a  baseline interpretation for each observational outcome.  It is clear that if any experiment reaches these forecasts level, we are going to learn a lot, no matter what we find, which is an ideal situation for an experiment to be.  In the event of a detection of either shape, measuring the scaling in the squeezed limit is an important distinguishing tool.

\subsection{Targets for the power spectrum}

The power spectrum of density fluctuations encodes a degenerate combination of the initial state and evolution of the primordial comoving horizon. In the context of inflationary cosmology, the evolution of the comoving horizon is fixed by the precise shape of the scalar field potential. Measuring the first two coefficients in a logarithmic expansion of the power spectrum, the spectral index $n_s$ and running $\alpha_s$, provides constraints on the inflaton potential. For example, the simplest single-field models of inflation would be ruled out by a measurement of significant running. Ultra-precise measurements of $n_s$ and $\alpha_s$ could greatly constrain the model-space of inflationary cosmology~\footnote{The utility of this exercise is arguably highly dependent on appropriate theoretical priors, as many models will be indistinguishable even within the ultimate cosmic variance limited error bars.}. 

%Studying the full phenomenology of inflationary models may require going beyond a simple parameterization, including for example oscillations in the power spectrum as suggested by some string theory motivated models~\cite{McAllister:2014mpa}. Oscillatory features in the power spectrum are also sourced by features in the inflaton potential~\cite{Adams:2001vc} and modifications of the vacuum state~(e.g. \cite{Easther:2001fi,Albrecht:2014aga}). 
Access to pre-inflationary initial conditions imprinted in the two-point function at the largest scales is possible when there is just-enough inflation. A host of ideas including an initial period of fast-roll~\cite{Hazra:2014jka,Pedro:2013pba}, excited states~\cite{Cicoli:2014bja,Lello:2013mfa}, and connections to the eternally inflating multiverse~\cite{Bousso:2013uia, Bousso:2014jca} have recently been invoked to explain the anomalously low power at $\ell \lesssim 30$~\footnote{Further motivation to study novel phenomena at large scales arises from a tension between the tensor power claimed to be observed by BICEP2 and the CMB temperature power spectrum, e.g.~\cite{Miranda:2014wga}.}.  Future LSS may provide improved constraints on the power on large scales~\cite{Bousso:2013uia}. In addition, an important exercise is determining how distinguishable all of these scenarios are by incorporating information beyond the two-point function (e.g. \cite{Holman:2007na,Adshead:2011jq}).  

Another signature of significant theoretical interest are oscillations in the power spectrum, bispectrum, and beyond.    This is motivated by the symmetry structure of string theory along axion directions in field space, e.g. as an auxiliary signature of axion monodromy inflation  \cite{MonodromyI}, as well as from the point of view of weakly broken discrete shift symmetries in low energy effective field theory \cite{EFToscillations}.  
The oscillatory features have a model-dependent amplitude which is exponentially sensitive to couplings in the theory, and may be undetectably small, but there are interesting theoretical thresholds in simple examples \cite{FMSW}.  In particular, in the case of high-scale inflation there are bounds on the coupling and size of extra dimensions in string theory, which translate into an interesting lower bound on the size of oscillations in some simple cases despite the exponential suppression.\footnote{With further assumptions about initial conditions, such as the possibility of tunneling from an oscillation-induced metastable minimum, one can deduce additional novel theoretical thresholds.}      

In the single-field version of axion monodromy inflation -- or any similar mechanism exhibiting a softly broken discrete shift symmetry --  one finds a potential for the canonically normalized inflation field $\phi$ of the form
\bea\label{genpot} 
V(\phi) &=& V_{0}(\phi) + \Lambda(\phi)^4 \cos[a(\phi)] \\ \nonumber
 &\simeq& V_0(\phi) + \Lambda(\phi)^4 \cos\Biggl[\frac{\phi_k}{f_0} \times \Biggl(1+\phi_{\star}\frac{df}{d\phi}\Bigr|_{\phi_{\star}}\left(\frac{\phi-\phi_{\star}}{\phi_{\star}}\right)+\frac{1}{2}\phi_{\star}^2\frac{d^2f}{d\phi^2}\Bigr|_{\phi_{\star}}\left(\frac{\phi-\phi_{\star}}{\phi_{\star}}\right)^2+\dots\Biggr)^{-1}\Biggr].\\ \nonumber
\eea
Here $a(\phi)$ is the underlying periodic axion variable, which in general is a nonlinear function of the canonical inflaton $\phi$.     
%We have also allowed for $\phi$-dependence in the amplitude of the oscillations.  
Nontrivial dependence $a(\phi)$ -- at a level crucial to include in the analysis -- is generic and can have multiple underlying causes, including back-reaction of the inflationary energy on other degrees of freedom, as well as loop effects derived from the weak explicit breaking of the discrete shift symmetry in $V_0(\phi)$.  The former is generally independent of the latter, leading to a wide parameter range of interest for searches \cite{FMSW}.     

The resulting power spectrum depends on the wavenumber $k$ of the scalar perturbations via a nontrivial function of $\log(k/k_{\rm pivot})$.   By computing appropriately normalized overlaps of power spectra, one can show that two nontrivial orders in the slow roll expansion, as well as two nontrivial orders in the Taylor expansion in the second line of (\ref{genpot}) must be included in order to capture the drifting oscillatory signature if present in the data.

Previous analyses \cite{oscillationsdata}\ for certain patterns of
non-drifting oscillations have led to bounds of roughly $\Lambda^4 < 10^{-3}
\sqrt{f_0/M_p}$.  The range of periods of interest includes $10^{-4}\le
f_0/M_p\le 1$, with lower values theoretically possible, and is particularly
sensitive to the ultraviolet theory.  This analysis including the parameters
associated with the drift in the period of oscillations is period is currently starting to be carried out in the CMB.  Resonant non-Gaussianity \cite{resonantNG}\ is another related signature of interest, requiring a similarly precise analysis of the bispectrum and higher point correlators\cite{Munchmyeretal2014}.

\section{Observational status and prospects for $\fnlloc$}\label{sec:local}
%Add a simple mode counting argument
%Currently planned survey (Anze): 1 page
%Array with forecasts?
Local non-Gaussianity is the best-studied inflationary signature for future surveys.  Having emphasized the need for sensitivity to $|\fnlloc| <1$, we now discuss the prospects for achieving this level of sensitivity in near-term surveys.

\subsection{Measurement status of primordial non-Gaussianity }

%To date, LSS has only been competitive with the CMB in the measurement of
%$\fnlloc$, through the scale dependent bias found by~\cite{Dalal:2007cu}.  In
%the presence of local non-gaussianity, the linear bias of tracers becomes
%\beq
%\delta_h(k)= \Big( b_1+ \frac{3 \Omega_m H_0^2}{c^2 k^2 T(k) D(z)} \fnlloc \frac{\partial \log n}{\partial \log \sigma_8} \Big) \delta(k)
%\eeq
%where $\delta_h$ is the tracer (halo) density contrast, $\delta$ is the matter density construct and $n$ is the number density of halos.

To date, LSS has only been competitive with the CMB in the measurement of
$\fnlloc$ and $\gnl$ through the scale dependent bias found by~\cite{Dalal:2007cu}.  In
the presence of $\fnlloc$-type non-gaussianity, the linear bias of tracers becomes
\beq
\label{eq:halobk}
\delta_h(k)= \Big( b_1+ \frac{3 \Omega_m H_0^2}{c^2 k^2 T(k) D(z)} \fnlloc \frac{\partial \log n}{\partial \log \sigma_8}(M) \Big) \delta(k)
\eeq
where $\delta_h$ is the tracer (halo) density contrast, $\delta$ is the matter density construct and $n$ is the number density of halos. Intuitively, the non-Gaussian correction to the bias can be understood as reflecting the coupling between the short-scale modes that form halos and long-wavelength fluctuations in the primordial potential (which is related to the density field via a factor of $k^2T(k)$). The scaling of the $\fnlloc$-dependent term approaches $1/k^2$ at low-$k$ making it a dramatic signature of non-Gaussian primordial statistics that, if detected, would rule out single-clock inflationary models \cite{Creminelli:2011rh}. 

The form of the non-Gaussian bias in Eq.~(\ref{eq:halobk}) has been repeatedly confirmed by N-body simulations (e.g. \cite{Pillepich:2008ka, Grossi:2009an, Scoccimarro:2011pz}). And, other local-type non-Gaussianities (e.g. $\gnl$ or scale-dependent $\fnlloc$ shapes) have been shown to generate corrections to the halo bias with the same $k$-dependence but with coefficients that have a different dependence on halo mass $M$  \cite{2010PhRvD..81b3006D, Shandera:2010ei, Smith:2011ub, Adhikari:2014xua}. 

Slosar et al \cite{Slosar:2008hx} produced the first constraints on $\fnlloc$
from scale-dependent halo bias shortly after it was identified as a signature
of primordial non-Gaussanity \cite{Dalal:2007cu}. They \cite{Slosar:2008hx} found $-29 < f_{NL}^\mathrm{loc} < +70$ (at $95\%$ confidence) from the clustering statistics of a variety of biased tracers: photometric luminous red galaxies (LRGs) from SDSS Data Release 6 (DR6), spectroscopic LRGs from SDSS DR4, photometric quasars from SDSS DR6, and NVSS radio galaxies (in cross-correlation with the CMB). The constraints in this analysis are dominated by data from quasars (which are highly biased and measured across large volumes) and quasar clustering alone yields $-69 <\fnlloc < 55$ (at $95\%$ confidence). These bounds on $\fnlloc$ were competitive with the contemporary CMB bispectrum constraints from WMAP 5 ($-9< \fnlloc < 111$) \cite{Komatsu:2008hk}.

Scale-dependent halo bias has tremendous potential to detect local-type primordial non-Gaussianity, but, as recognized already in \cite{Slosar:2008hx}, instrumental, observational, and astrophysical systematics can generate spurious power on large-scales that mimics the effect. Indeed, a number of subsequent analyses have found excess power in the large-scale quasar and galaxy clustering that could be interpreted as evidence for primordial non-Gaussianity, but is generally assumed to be a systematic effect \cite{Afshordi:2008ru,Xia:2010pe,Xia:2011hj, Ross2011,Ross:2012sx, Pullen:2012rd, Giannantonio:2013uqa,Ho:2013lda, Agarwal:2013qta,  Karagiannis:2013xea, Leistedt:2014zqa, Leistedt:2014wia}. Dust extinction, stellar contamination, variations in seeing and sky brightness, for instance, can modulate the observed number density of sources on very large scales and these large-scale power modulations are difficult to separate from large-scale modulations in the source density due to local primordial non-Gaussianity (see, for instance \cite{Huterer:2012zs, Pullen:2012rd, Agarwal:2013ajb, Giannantonio:2013uqa}).

Current analyses treat systematics by restricting to data products that are well-understood, by modeling systematic effects and correcting the measured power spectra or correlation functions, and by projecting out modes that appear correlated with templates of known systematics (e.g. \cite{Ross:2012sx, Agarwal:2013ajb, Giannantonio:2013uqa,Ho:2013lda, Agarwal:2013qta, Leistedt:2014zqa}). These techniques strengthen the confidence in current bounds on primordial non-Gaussianity, but systematics remain an important limitation to constraints on primordial non-Gaussianity from scale-dependent bias. 

 At present, the most stringent constraints on primordial non-Gaussianity from large-scale structure are from photometric quasars from SDSS DR8 by Leistedt et al \cite{Leistedt:2014zqa}. This work treats systematics by projecting out modes of the quasar field that exhibit significant correlation with template maps, and products of template maps, of possible systematic effects in SDSS. The final constraints from this analysis are $-49 < f_{NL}^\mathrm{loc} < 31$ (at $95\%$ confidence). These constraints on $f_{NL}^\mathrm{loc}$ are only marginally tighter than the initial constraints from all probes given in Slosar et al \cite{Slosar:2008hx}, but significantly tighter than the first constraints from quasars alone (quoted above). More importantly, the current limits should be more robust against observational systematics. 
 
 Finally, a number of analyses have used scale-dependent bias to constrain models of non-Gaussianity beyond the usual local ansatz. The leading contributions to the non-Gaussian bias can be more generally modeled as \cite{Agarwal:2013qta}
\begin{equation}
	\Delta b \left( M,k,z,{\mathcal A}_{\rm NL},\alpha \right) \propto [b_{1}(M,z) - p] \frac{{\mathcal A}_{\rm NL} \left( b_{1}(M,z) \right)}{k^{\alpha}}
\label{eq:genNGbias}
\end{equation}
where $M$ is the mass of the tracer and $p$ different from one allows for tracers whose bias may depend on the merger history of the host halo \cite{Slosar:2008hx}. This parameterization captures, for example, the effects of allowing a cubic $g_{\rm NL}$ term in the local ansatz \cite{Smith:2011ub}, two fields with different power spectra contributing to the fluctuations \cite{Shandera:2010ei}, the quasi-single field models \cite{Norena:2012yi, Sefusatti:2012ye}, general initial states for the inflationary fluctuations \cite{Ganc:2012ae, Agullo:2012cs}, a scale-dependent local ansatz \cite{Becker:2010hx, Shandera:2010ei}, and arbitrary scale dependence of the bias \cite{Leistedt:2014zqa}. When multiple fields contribute to the fluctuations, $\tau_{\rm NL}\geq(\frac{6}{5}f_{\rm NL}^{\rm loc})^2$ and the non-Gaussian bias is in addition stochastic \cite{Tseliakhovich:2010kf, Baumann:2012bc}. The more general form in Eq.(\ref{eq:genNGbias}) is not well constrained with current data but several authors have considered more specific extensions of the local scenario \cite{Leistedt:2014zqa, Giannantonio:2013uqa, Tseliakhovich:2010kf, Agarwal:2013qta}. Scale-dependent bias from SDSS DR8 photometric quasars limits $-2.7 < g_{\rm NL}^{\rm loc} \times 10^{-5} < 1.9$ or $-105 < \fnlloc < 72$ and $-4 < g_{\rm NL}^{\rm loc} \times 10^5 < 4.9 $ when $f_{NL}^\mathrm{loc}$ and $g_{NL}$ are analyzed jointly (all intervals are $95\%$ confidence) \cite{Leistedt:2014zqa}. Reference \cite{Leistedt:2014zqa} constrained departures from $\alpha=2$ by $b(k)\propto k^{-2+\delta}$ and obtain constraints\footnote{Those authors used a notation $\mathcal{A}=\tilde{f}_{\rm NL}$ and $\delta =n_{f_{\rm NL}}$. However, that notation for the departure from the local ansatz may be confusing since introducing a scale dependence in the amplitude of the bias does not correspond to introducing a scale-dependent $f_{\rm NL}$ in the local ansatz. A majority of the literature uses $n_{f_{\rm NL}}$ to mean a scale dependent amplitude of the local ansatz, as $\zeta(x)=\zeta_{\rm Gauss.}(x)+\frac{3}{5}f_{\rm NL}\star\zeta(x)^2$, promoting the multiplication to a convolution.} $-45.5{\rm exp}(3.7\delta)<\mathcal{A}<34.4{\rm exp}(3.3\delta)$. Finally, a different approach was taken in \cite{Becker:2012yr}, who forecast constraints on a scale-dependent local model where $f_{\rm NL}$ is a free parameter in several $k$-space bins. Forecasts for improved constraints on multi-field models with scale-dependence can be found in \cite{raccetal14}.
 
 %Finally, a number of analyses have used scale-dependent bias to constrain more general models of local non-Gaussianity. For instance, by including a cubic $g_{NL}$-term, allowing the $k-$dependence of the non-Gaussian bias to be a general power law, or allowing contributions from a second field so that $\tau_{NL} \ge (\frac{6}{5}\fnlloc)^2$ so that the non-Gaussian bias is stochastic with respect to the matter density \cite{Tseliakhovich:2010kf, Giannantonio:2013uqa, Agarwal:2013qta, Leistedt:2014zqa}. Scale-dependent bias from SDSS DR8 photometric quasars limits $-2.7 < g_{\rm NL}^{\rm loc} \times 10^{-5} < 1.9$ or $-105 < \fnlloc < 72$ and $-4 < g_{\rm NL}^{\rm loc} \times 10^5 < 4.9 $ when $f_{NL}^\mathrm{loc}$ and $g_{NL}$ are analyzed jointly (all intervals are $95\%$ confidence) \cite{Leistedt:2014zqa}. When introducing a running parameter $n_{\fNL}$ defined by $b(k) \propto k^{-2+n_{\fNL}}$ with a non-Gaussian amplitude $\tilde{f}_{\rm NL}$, reference \cite{Leistedt:2014zqa} find those parameters to be limited to $-45.5 \exp(3.7 n_{\fNL}) < \tilde{f}_{\rm NL} < 34.4 \exp(3.3n_{\fNL})$  at $95\%$ confidence. 

%Forecasts for PS and BS when available

\subsection{Relevant planned surveys}

There are currently several large-scale structure surveys -- either currently
underway, funded or proposed -- that have the potential to constrain
the parameters and physics of inflation. They fall into three broad
classes: 
\begin{enumerate}
\item \textbf{Spectroscopic galaxy surveys} measure redshifts of
  targeted galaxies on the sky. These objects act as tracers of the
  underlying curvature fluctuations. 
\item \textbf{Photometric galaxy surveys} measure the luminosity of
  galaxies on the sky in several wide-bands. Galaxy type and a rough
  redshift can be deduced from this information. The information about
  radial distance is mostly lost, but this is compensated by many more
  objects observed.
\item \textbf{21-cm galaxy surveys} measure the integrated emission
  from the 21-cm hydrogen spin-flip transition. Due to foreground removal
  processes, the low-$k$ radial fluctuation modes are lost.
\end{enumerate}

In Table \ref{tab:surveys} we list several important LSS experiments happening
towards the end of this decade that could have influential impact on our
understanding of inflation.  The Table necessarily oversimplifies aspects of
each experiment and therefore numerical values should be compared with
care. For the same reason we refrain from making any specific forecasts.

In general, galaxy clustering power spectrum and bispectrum
measurements will provide measurements of the $f_{\rm NL}$ parameters
with accuracy up to around $1$ for the local shape and somewhat worse for
other bi-spectrum shapes. With sufficient modeling, the galaxy power
spectrum and Lyman-$\alpha$ flux power spectrum can provide
information on the spectral index and its running and constrain possible
sharp features or oscillations in the primordial power spectrum.  They
will also be able to further improve limits on curvature parameter
$\Omega_k$ and the fraction of isocurvature fluctuations.  Weak
lensing measurements will be able to provide information on the same
parameters with very different systematics: they probe the dark-matter
directly which simplifies theoretical treatments, but have fewer modes
and challenging-to-control observational systematics. Note also that none of the LSS
experiments will be able to provide meaningful information
on the presence of primordial tensor modes.

\begin{table}
  \centering
  \resizebox{16.75cm}{!} {
  \begin{tabular}{|p{3.0cm}|C{3.0cm}|C{3.0cm}|C{3.3cm}|C{3.3cm}|C{2.5cm}|}
        \hline & {LSST} & DESI & Euclid & SPHEREx & CHIME \\
    \hline
    \hline
 Survey type & photo  & spectro & photo+spectro  & low-res spectro & 21-cm \\
\hline
Ground or space & ground & ground & space & space & ground \\
\hline
Previous surveys & CFHTLS, DES, HSC & BOSS, eBOSS, PFS & no direct precursor & PRIMUS, COMBO-17, COSMOS& GBT HIM \\
\hline
Survey start &   2020 &  2020 &  2018 &  2020 &  2016 \\
%\hline
%Etendue$^a$ [m$^2.\deg ^2$] &  320 ($\times \sim 0.73$) &  70.2 ($\times \sim 0.9$) &  1.25 ($\sim \times 0.73$)  &  3.1 &  N/A \\ 
\hline
Redshift-range & $z<3$ (1\% sources above 3) & $z<1.4$, $2<z<3.5$ (Lya) & $z<3$ & $z<1.5$ & $0.75<z<2.5$\\
\hline
Survey area [$\deg^2$] & 20k & 14k & 15k & 40k & 20k \\
\hline
Approximate number of objects & $2\times 10^9$ (WL sources) & 22$\times10^6$ gal., $\sim 2.4\times 10^5$ QSOs &
$40\times 10^6$ redshifts, $1.5 \times 10^9$ photo-$z$s & $15\times 10^9$ pixels &  $10^7$ pixels \\
\hline
Galaxy clustering & \dmark$^\diamond$ & \cmark & \cmark & \cmark & \cmark\\
Weak lensing &\cmark&&\cmark&& \cmark\\
RSD              & & \cmark & \cmark & \dmark & \dmark\\
Multi-tracer &\dmark &\dmark & \dmark & \cmark & \\
%\hline
%Approximate price-tag [M\$] & 600 & 70 & 600 & 125 & 30 \\
%Funding status & funded & passed DOE CD1 & funded & proposed & funded \\
\hline
\end{tabular}}
  \caption{A selection of currently funded or planned surveys. Other important surveys not included in the table are PFS, JPAS, PAU, EMU. Relevant survey links \href{http://www.lsst.org/}{[LSST]},\href{http://desi.lbl.gov/}{[DESI]},\href{http://sci.esa.int/euclid/}{[Euclid]},
\href{http://http://chime.phas.ubc.ca/}{[UBC]},\href{http://http://sumire.ipmu.jp/pfs/}{[PFS]},
\href{http://j-pas.org/}{[JPAS]},\href{http://http://www.ieec.cat/project/pau-physics-of-the-accelerating-universe/}{[PAU]}, \href{http://www.atnf.csiro.au/people/Ray.Norris/emu/index.html}{[EMU]}. $^\diamond$Galaxy clustering is possible, but very strong radial degradation.}

%%Notes: Etendue is the product of (effective mirror size) with field of view and measures relative survey speed. %Due to different observing conditions one should not compare numbers for space and ground missions. %Mirror-size losses are estimated based on 10\% loss per reflection. 
%$Currently funded or planned surveys. %$Notes: $^a$Etendue is the product of
 %(effective mirror size) with field of view and measures relative survey speed. 
%Due to different observing conditions one should not compare numbers for space 
%and ground missions. Mirror-size losses are estimated based on 10\% loss per reflection.
%$^b$Angular galaxy clustering is possible, but serious radial
%degradation. \dragan{Some more vertical centering necessary to line up first
 % column with later ones. Also, I find the half- and full-circle notation a bit weird -
 % single and double check marks or something would work better maybe...}
 % }
\label{tab:surveys}
\end{table}

%%%%%%%%%%%%%%%%%%%%%%%%%%%%%%%%%%%%%%%%%%%%%%%%%%%%%%%%%%%%%%%%%%%%%%%%
\subsection{What would an ideal survey for constraining $f_{\rm NL}^{\rm loc}$ with the power spectrum look like?}
%%%%%%%%%%%%%%%%%%%%%%%%%%%%%%%%%%%%%%%%%%%%%%%%%%%%%%%%%%%%%%%%%%%%%%%%

Having considered the expected constraints on primordial non-Gaussianity from
upcoming surveys, it is instructive to ask what would be an ideal experiment
to constrain it.  We will focus here on primordial non-Gaussianity of the {\it
  local} type as constrained by scale-dependent bias in the galaxy {\it power
  spectrum}.  Based on Section \ref{sec:theory}, let us take $\sigma(\fnlloc)
\sim 1$ as a target for the ideal survey.  The results in this section will be
largely based on \cite{fnlsurvey}, but see also other recent works,
e.g.~\cite{ferrsmith14,raccetal14}.  The forecast numbers are necessarily very
rough as they depend on the various survey parameters.  We quote them to give an
approximate quantitative sense of what is required of a survey and refer to
\cite{fnlsurvey} for more details.

The crucial general consideration for optimizing a survey to measure $\fnlloc$
is that the scale-dependent bias scales as $\Delta b \propto k^{-2}$ (at small $k$),
so that the signal is dominated by the largest scales probed by the survey.
Turning now to specific survey properties,
the ideal {\it number density} of galaxies is such that the shot noise is comparable to or a bit smaller
than the clustering signal, i.e.~$\bar{n} P_g(k) \gtrsim 1$. Here, $P_g(k)$ is the galaxy power spectrum, which is
to be evaluated at the typical wave vector $k$ important for $\fnlloc$ (say $k
\sim 0.001\, h^{-1}$Mpc).
This leads to an approximate requirement $\bar{n} \approx$ few $ \times 10^{-4}\, (h^{-1}$Mpc$)^{-3}$,
depending on redshift and the bias of the galaxy sample.
Since the $\fnlloc$ signal is dominated by very large scales, cosmic variance (i.e.~the availability
of only a small number of modes) is a major limitation. This can partially be undone by analyzing multiple galaxy samples
with different biases. Using this multi-tracer technique the information per unit volume can be significantly improved
relative to the single-tracer case. However, to obtain a strong gain a much larger total number density of galaxies
is required, say $\bar{n} \gtrsim$ few $\times 10^{-3}\,(h^{-1}$Mpc$)^{-3}$.

Next, because of the limited number of modes on the largest scales, the requirement on survey {\it volume} is
extremely stringent. To obtain $\sigma(\fnlloc) \approx 1$ using a single-tracer analysis,
volumes $V \sim $ 
%DH: Fixed volume units from Gpc^{-3} to Gpc^3
few $\times \, 100\, (h^{-1} $Gpc$)^{3}$ are needed. If the multi-tracer technique
can be employed (requiring a much larger number density as discussed above),
the volume requirement can be loosened to, say, $V \sim 100 \, (h^{-1} $Gpc$)^{3}$.

\begin{figure*}[ht!]
\includegraphics[width=0.48\textwidth]{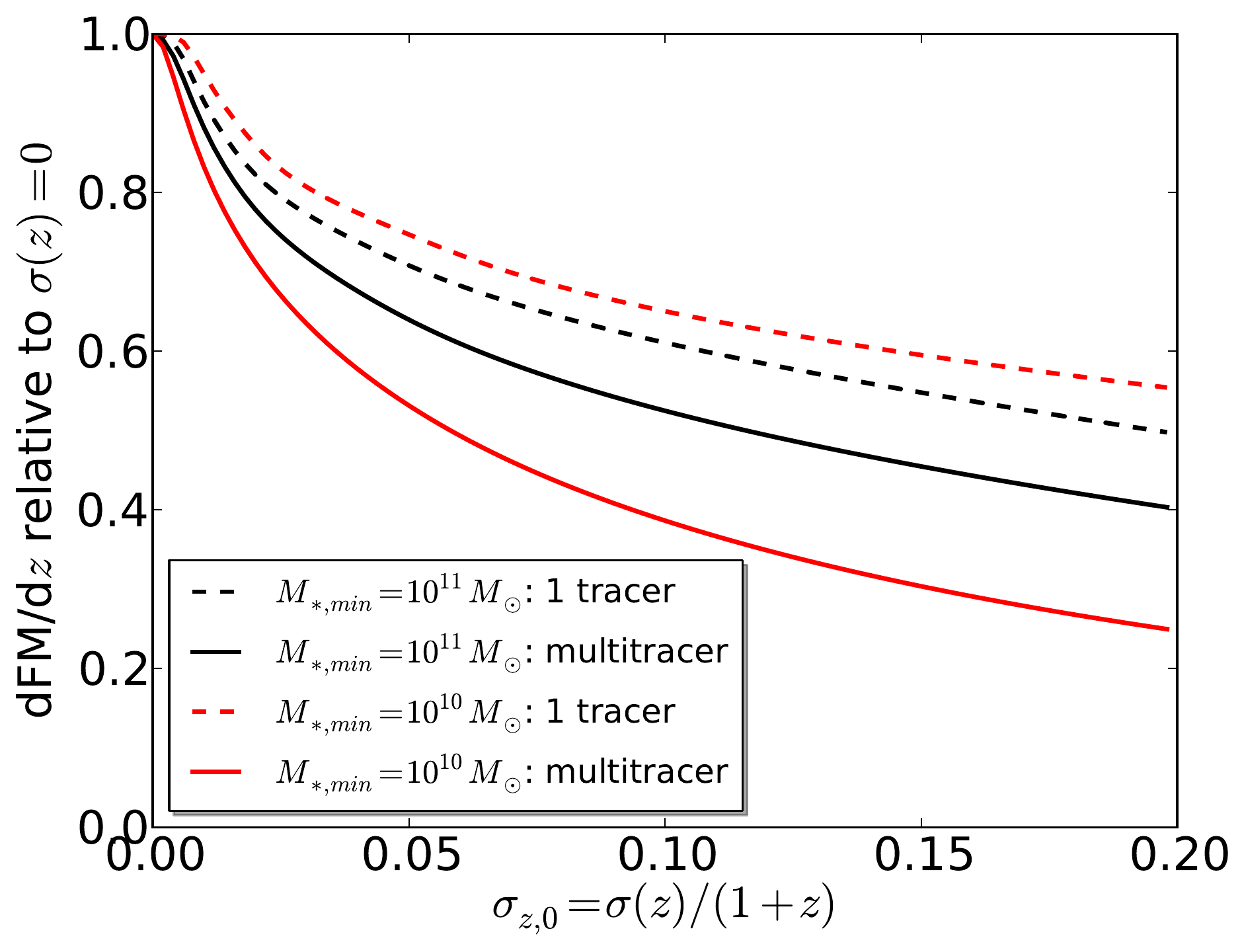}
\includegraphics[width=0.48\textwidth]{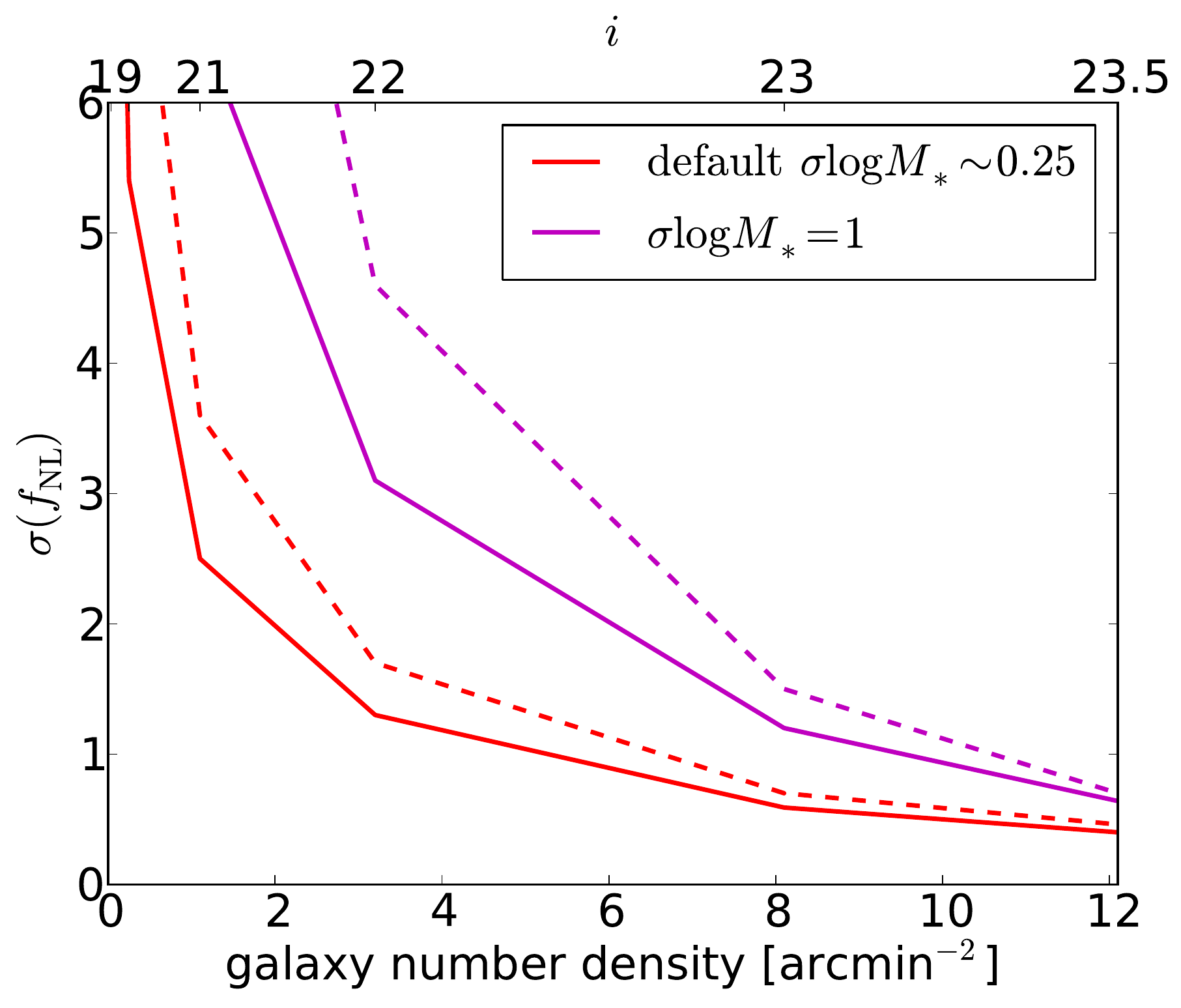}
\caption{{\it Left:} Fisher information on $f_{\rm NL}$ (proportional to $1/\sigma^2(f_{\rm NL})$)
relative to that of a survey with perfect (spectroscopic) redshifts, as a function of redshift scatter.
Solid curves assume use of the multi-tracer technique and dashed assume a single tracer.
We show results for moderate and high number density surveys, represented by stellar mass cuts
$M_* > 10^{11} M_\odot$ and $M_* > 10^{10} M_\odot$, respectively.
{\it Right:} Single-(dashed) and multi-(solid) tracer uncertainties on $f_{\rm NL}$
for a toy model of a multi-band imaging survey, as a function
of an $i$-band magnitude cut (and corresponding number density on the lower horizontal axis).
We assume a sky coverage $f_{\rm sky}= 0.75$.
Stellar mass is used as a proxy for host halo mass (and therefore bias) and we show results for two different values of the stellar mass scatter
relative to the mean stellar mass - halo mass relation.
For both figures, we refer to the text and \cite{fnlsurvey} for details.
}
\label{fig:idealsurvey}
\end{figure*}

The dependence of the signal on large scales also has a positive side: the required redshift accuracy is not
very high. In general, any scatter in the redshift estimator will lead to a smoothing of the clustering signal
in the radial direction, with the comoving smoothing scale proportional to the redshift scatter,
$d_s = \sigma_z/H(z)$, where $H(z)$ is the Hubble parameter. In order for this smoothing to not
destroy the $\fnlloc$ signal, the smoothing scale needs to be smaller than the scale of the modes of interest,
$d_s \, k < 1$. In practice, this means that as long as $\sigma_z < 0.1 (1 + z)$, most of the constraining power
on $\fnlloc$ is conserved (see Figure \ref{fig:idealsurvey}, left panel).

Finally, in addition to a galaxy's redshift, we wish to be able to measure a galaxy property that strongly correlates with
host halo mass (which in turn mainly determines clustering bias). In the multi-tracer case, this would make it
possible to subdivide a sample into different subsamples with as much difference in bias as possible (as this boosts
the $\fnlloc$ significance). Even in the single-tracer scenario, having some type of mass proxy will make it possible
to choose the single sample cleverly. For instance, a sample with very high number density at some redshift
will typically have a galaxy bias not very different from unity and therefore
a weak scale-dependent bias signature
in the presence of primordial non-Gaussianity. With mass information available, a subsample of galaxies living in more massive
halos could be selected, thus boosting the bias and therefore the $\fnlloc$ signal (of course one has to be cautious not
to select so small a subsample that the increase in shot noise cancels the gains from the increase in bias).

Based on the above, we can now ask what type of galaxy survey is optimal for constraining $\fnlloc$.
Interestingly, an optimal $\fnlloc$ survey would look very different than the currently prevalent type of
cosmological galaxy clustering survey. Such surveys are typically optimized for baryon acoustic oscillations (BAO),
and other physics with a large signal down to small scales.
This naturally leads to surveys with {\it spectroscopic} redshifts (to measure
clustering down to $k_{\rm max} = 0.1 - 0.2 h/$Mpc).
On the contrary, the mild redshift accuracy requirement and stringent requirements on survey volume
and number density strongly suggest that a spectroscopic survey is not optimal for $\fnlloc$, as a lot of time
would be spent achieving a better-than-needed redshift accuracy, limiting the total number of galaxies.
Instead, it appears that \emph{a large area (ideally full-sky), multi-band, imaging survey would be ideally suited for constraining primordial non-Gaussianity from scale-dependent halo bias.} The redshifts would thus be photometric redshifts or, in the case of a  survey with a large number of narrow bands, low-resolution spectroscopic redshifts. Typically, this will mean a very high number density 
at low redshift, with a decrease towards higher redshift. Thus, at low redshift, the multi-tracer technique can
be applied to boost the $\fnlloc$ constraint, whereas at the higher redshift end of the survey
using a single tracer is close to optimal and there is no multi-tracer benefit.

Based on a toy model for such an imaging survey, \cite{fnlsurvey} found that a full-sky (in practice $f_{\rm sky} = 0.75$) survey collecting
a sample complete up to magnitude $i \sim 22 - 23$, corresponding to a total
angular number density $\bar{n}_A \gtrsim 5$ arcmin$^{-2}$, would be able to
obtain $\sigma(\fnlloc) = 1$ (see Figure \ref{fig:idealsurvey}, right panel). This assumes that the survey measures stellar mass with a typical scatter relative to the mean stellar mass - halo mass relation of $\sigma(\log_{10}M_*) \lesssim 0.25$,  or a different halo mass proxy with comparable scatter. It also assumes redshifts are measured better than $\sigma_z = 0.1 (1 +  z)$. Relaxing those constraints would weaken the $f_{\rm NL}$ bound.

\section{Beyond the Halo Power Spectrum} \label{sec:bispec}

As emphasized in Section~\ref{sec:theory}, not all theoretical targets will leave clean signatures in the scale dependent bias of halos.  Equilateral non-Gaussianity is one particularly interesting observable that requires a measurement of the halo bispectrum (for example) to be probed by large scale structure.  In this section we will discuss the other probes and their status as tools for testing inflation.

\subsection{State of Bispectrum Observations}

Compared to extensive studies of galaxy clustering measurements at the
two--point level, progress on the three--point level has been much
slower. After a few early detections of the galaxy
bispectrum~\cite{Scoccimarro01,Verde02,Scoccimarro04}, as well as measurements
of the galaxy three--point correlation function in several surveys the
2dFGRS~\cite{Jing04,Gaztanaga05}, SDSS~\cite{Marin11,McBride11},
WiggleZ~\cite{Marin13}, \cite{GilMarin14} present a robust measurement and
cosmological analysis~\cite{GilMarin14b} of the galaxy bispectrum monopole
(i.e., angle--averaged with respect to the line of sight direction) using the
SDSS CMASS galaxy sample. This is the largest data set for which a bispectrum
measurement has been obtained so far, and provides unprecedented
signal--to--noise. The analysis finds agreement between the measured galaxy
bispectrum and bispectra derived mock catalogs few percent--level up to
$k_{\mathrm{max}}\sim 0.2 h/\mathrm{Mpc}$ and at $\langle z\rangle = 0.57$. On
large angular scales, the measurement is affected by approximations in the
modeling of survey geometry effects, which are required due to computational
complexity of the bispectrum estimator in presence of a survey mask, limiting
the analysis to $k_{\mathrm{min}} = 0.05\, h\,\mathrm{Mpc}^{-1} \approx 3
k_{\mathrm F}$ where $k_{\mathrm F}$ is the fundamental mode of the survey. The combination of power spectrum and bispectrum measurements is then used to determine the amplitude of mass 
fluctuations $\sigma_8$ and the linear growth rate $f\equiv d\ln D/d\ln a$ from clustering data alone for the first time, and the results are in broad agreement with the CMB-inferred values.

\subsection{State of Bispectrum Forecasts} 
Forecasts for the constraining power of the galaxy bispectrum on primordial
non-Gaussianity are few, and they typically lack the detailed modeling found
in power spectrum forecasts (e.g.~\cite{Giannantonio12}). Based on idealized
survey geometry assumptions and ignoring the redshift evolution,
\cite{Scoccimarro04} estimate that an all-sky redshift survey with
$\bar{n}\sim 3\times 10^{-3} \, (h/\mathrm{Mpc})^3$ up to $z\sim1$ will probe
$\sigma(\fnlloc)\sim1$. Using more detailed modeling, \cite{Sefusatti07} find
$\sigma(\fnlloc) = 3.6$ and $\sigma(\fnlequil) = 42$ for a $100\,
(h^{-1}\mathrm{Gpc})^3$ survey and sources at $1 \lesssim z\lesssim 2$ with
$\bar{n}\sim 10^{-3} \, (h\,\mathrm{Mpc}^{-1})^3$ (c.f. their Table 1 for forecasts
based other survey parameters). Note that both of these studies predate the
discovery of primordial non-Gaussianity induced scale-dependent galaxy bias,
and thus do not include this key signature of primordial
non-Gaussianity in the galaxy distribution.

The bispectrum of LSS has a vast potential to directly probe interactions in
the early Universe. This is especially true in regards to equilateral and
orthogonal shapes that, unlike local non-Gaussianity, do not lead to an upturn
in the bias on large scales; these shapes can in principle be probed by the three-point function. As an added benefit, the bispectrum provides more modes than the power spectrum since there are more triangle configurations than line-configurations in $k$-space.
However, this promise is hampered by the significant late time non-linearities
caused by gravitational clustering and intrinsic non-linearities of the tracer
population. Furthermore, the bispectrum has comparably large errors on large
scales, so that for sufficient signal-to-noise parameter inferences one needs to go into the weakly non-linear regime where loop corrections to the spectra are important.
We will first focus on the status of the modeling of the mundane contaminants present for Gaussian initial conditions and describe the modifications arising from primordial non-Gaussianity in the end.

\subsubsection{The Matter Bispectrum}
The matter bispectrum on large scales (small wavenumbers) is well understood in Standard Perturbation Theory \cite{Bernardeau:2002la} at tree level
\[B_\text{tree}(\vec k_1,\vec k_2,\vec k_3)=B_{211}(\vec k_1,\vec k_2,\vec k_3)=2F_2(\vec k_1,\vec k_2)P(k_1)P(k_2)+\text{2 cyc.}\]
Going to higher wave numbers requires one to consider the loop corrections to the above equation that are sensitive to the third and fourth order gravitational interaction kernels
\[B_\text{1-loop}=B_{211}+B_{411}+B_{321}^{I}+B_{321}^{II}+B_{222}\]
A different approach has been suggested by \cite{2001MNRAS.325.1312S,2012JCAP...02..047G}, who instead replace the exact $F_2$ kernel in the tree level expression by a parametrized form, whose free parameters are fit to simulations
\[F_2^\text{eff}(\vec k_1,\vec k_2)=\frac{5}{7}a(k_1)a(k_2)+\frac{1}{2}\mu b(k_1)b(k_2)\left(\frac{k_1}{k_2}+\frac{k_2}{k_1}\right)+\frac{2}{7}\mu^2 c(k_1)c(k_2)\ ,
\]
where in SPT we have $a,b,c=1$ and $\mu=\vec k_1 \cdot \vec k_2/k_1 k_2$.
A more formal approach is to consider parametrized source terms in the Euler equation that account for higher moments in the Vlasov hierarchy and higher order non-linearities. The bispectrum in this Effective Field Theory of Large Scale Structure approach has been recently computed in \cite{2014arXiv1406.4143A,2014arXiv1406.4135B}. The performance at redshift $z=0$ for a standard $\Lambda$CDM model is as follows: tree level SPT fails at $k\approx0.05 \ihMpc$, one-loop SPT fails at $k\approx0.1 \ihMpc$ and EFT can extend this range to $k\approx0.25 \ihMpc$.\\

Yet another promising approach is the halo model \cite{2002PhR...372....1C} that disentangles the dark matter correlations into the correlations between halo centers and the dark matter profiles around these centers. The bispectrum is then given by the sum of one-, two- and three halo terms
$B_\text{mmm}=B_\text{1h}+B_\text{2h}+B_\text{3h}$.
With perfect knowledge of halo-halo-halo correlations and the halo profile, this model should provide a perfect description of the matter bispectrum. Unfortunately the ingredients are not understood at the required level of accuracy yet. A big advantage of the halo model is it can be easily extended to describe galaxy clustering using a halo occupation framework, as we discuss in more detail below.

%%%%%%%%%%%%%%%%%%%%%%%%%%%%%%%%%%%%%%%%%%%%%%%%%%%%%%%%%%%%%%%
\subsubsection{The Halo Bispectrum}
Dark matter haloes are a specific subset of the total matter distribution that can be best understood by selecting regions that will collapse in the initially weakly non-Gaussian field (Lagrangian bias), and then following the gravitational motion of the center of mass of these regions (leading to Eulerian bias). Collapsing regions preferably populate high density regions, which is encoded in their clustering statistics being enhanced with respect to the matter by a constant bias factor on large scales. At late times the matter density field can be schematically written as \cite{2009JCAP...08..020M}
\[\delta_\text{h}=b_1 \delta+\frac{b_2}{2}\delta^2+b_{s^2} s^2+b_{\Delta} \Delta \delta+\ldots \ , \]
where $s^2$ is the square of the tidal tensor $s_{ij}=\left(\partial_i\partial_j/\partial^2-\delta_{ij}^\text{(K)}/3\right)\delta$.
In principle one can write down all the terms allowed by symmetry and their coefficients can be estimated from the evolution of a reasonable Lagrangian bias model.\\

An additional complication arising for discrete tracers is their stochasticity or sampling variance. For a Poisson sampling this leads to a constant contribution to the halo power spectrum and to a contribution of the form
\[B_\text{hhh,shot}=\frac{1}{\bar n^2}+\frac{3}{\bar n}P_\text{hh}\]
for the bispectrum, where $\bar n$ is the number density of the tracers. As shown in \cite{2013PhRvD..88h3507B} for the power spectrum, this model receives corrections due to small scale exclusion and non-linear biasing. These corrections have been understood for the power spectrum but not yet for the bispectrum. Finally we note that an accurate description of the halo bispectrum is a key ingredient for halo model based calculations of the matter or galaxy bispectra (or their cross-bispectra).

%%%%%%%%%%%%%%%%%%%%%%%%%%%%%%%%%%%%%%%%%%%%%%%%%%%%%%%%%%%%%%%
\subsubsection{Redshift Space Distortions}

Galaxy positions are obtained by measuring two angles and a redshift. The latter is a combination of the Hubble expansion and the motion of the galaxy in the local gravitational potential, also known as the peculiar velocity. The peculiar velocities distort the distance estimation based on the Hubble flow and need to be accounted for, since they are correlated with the density fluctuations that we are trying to estimate. This is a complicated endeavor for two point functions and even more so for the three point function.\\

The tree level bispectrum in redshift space can be expressed as \cite{2014arXiv1407.5668G,1995A&A...298..643H}
\begin{align*}
B(\vec k_1,\vec k_2,\vec k_3)=&D_\text{FoG}\left[Z_1(k_1)P(k_1)Z_1(k_2)P(k_2)Z_2(\vec k_1,\vec k_2)+\text{2 cyc.}\right]\\
Z_1(k_i)=&b_1+f \mu_i^2 \\
Z_2(\vec k_1,\vec k_2)=&b_1\left[F_2(\vec k_1,\vec k_2)+\frac{f\mu k}{2}\left(\frac{\mu_1}{k_1}+\frac{\mu_2}{k_2}\right)\right]+f\mu^2 G_2(\vec k_1,\vec k_2)\\
&+\frac{f^3\mu k}{2}\mu_1\mu_2\left(\frac{\mu_1}{k_1}+\frac{\mu_2}{k_2}\right)+\frac{b_2}{2}+\frac{b_{s^2}}{2}S_2(\vec k_1,\vec k_2)\, .
\end{align*}
Here $D_\text{FoG}$ is a Lorentzian or Gaussian accounting for the small scale velocity dispersion within halos, $G_2$ is the velocity divergence coupling kernel, and $S_2$ describes the tidal term $s^2$ in Fourier space.
The above formula has been extended by \cite{2014arXiv1407.5668G} employing effective $F_2$ and $G_2$ kernels mentioned above. The halo model provides a way to explicitly account for the occupation of halos by galaxies and to derive the resulting bispectrum \cite{2008PhRvD..78b3523S}.

%%%%%%%%%%%%%%%%%%%%%%%%%%%%%%%%%%%%%%%%%%%%%%%%%%%%%%%%%%%%%%%
\subsubsection{Loop corrections to the halo bispectrum}

In the matter distribution, non-Gaussianity imprints an explicit bispectrum at tree level and adds additional loop corrections to the power spectrum. For discrete tracers, the non-Gaussian nature of the initial conditions affects the spectra resulting from the sampling process. For local non-Gaussianity the local variance of the field is correlated with the long wavelength curvature fluctuations. Since the local abundance depends on the local variance, the halo abundance is correlated with the curvature fluctuations leading to an additional bias parameter.\\

The non-Gaussian correction to the tree level real space halo bispectrum can be schematically written as \cite{Baldauf:2010vn,2012MNRAS.425.2903S}
\[\Delta B_\text{hhh}=b_1^2 \Delta b_1 B_\text{mmm,G}+b_1^3 B_\text{mmm,nG}+b_1^2 \Delta b_2 P^2+\ldots\, ,\]
where $\Delta b_1$ and $\Delta b_2$ are the bias corrections arising from the explicit correlation between long wavelength curvature fluctuations and small--scale variance. Preliminary and idealistic estimates \cite{Baldauf:2010vn} show that the bispectrum can provide tighter constraints on local non-Gaussianity, especially for $b_1\approx 1$, for which $\Delta b_1=0$ and consequently no enhancement in the power spectrum.\\

For equilateral or orthogonal shapes the second order bias corrections relevant for the bispectrum have not yet been derived, so that predictions and forecasts to date are based on the non-Gaussian $b_1$ multiplying the Gaussian matter bispectrum and the Gaussian $b_1$ multiplying the non-Gaussian matter bispectrum. They are missing the correction from the non-Gaussian second order bias $\Delta b_2$, multiplying two power spectra, which will probably increase the signal and change the shape of the template. For general shapes the correction to $b_2$ will itself have a shape itself.

\subsubsection{Additional Effects}

The above discussion skipped a number of observational effects that affect both the power spectrum and the bispectrum. The first one is that the survey mask --complete understanding o the survey geometry, especially important for harmonic-space measurements. The survey mask is  difficult to account for already in the power spectrum already \cite{2014MNRAS.443.1065B} and might be even more difficult in the bispectrum. On top of this are the relativistic effects discussed in the previous section. On large scales one has to give up the flat sky approximation as well and consider finite angle effects \cite{2012JCAP...10..025B,2014JCAP...08..022R} in conjunction with the relativistic effects. Given the complications in modelling and measuring the bispectrum, one might ask oneself whether it is necessary to go through the complication of measuring the full bispectrum in Fourier space and then applying the estimator on the measured bispectrum, rather than making the measurements in real space and applying a suitably defined real-space estimator of non-Gaussianity.

%Bispectrum theory and forecast status. 0.5 page
%Bispectrum observational status 0.5 page
\subsection{Mass function \& other probes}

While the bispectrum contains a great deal of information, no single statistic can completely characterize non-Gaussian fluctuations. Higher order correlations contain additional information. And, although we currently have good reason to believe that the bispectrum may be the most important observable for distinguishing models of primordial physics, it is important to consider additional observational tests that would confirm and strenghten inferences made from the power spectrum and the bispectrum. Attempting to directly measure or constrain the trispectrum and even higher-order correlations is a possible route, but one that becomes more difficult at each order as the possible functions of momenta become more complicated.

Fortunately, many large scale structure observables depend on sums of integrated (or partially integrated) correlations. Although such observables may be primarily sensitive to the amplitude of the bispectrum (assuming higher moments fall off in amplitude at each order) they also contain information beyond the bispectrum. The halo mass function is a simple example, where the difference between Gaussian and non-Gaussian cosmologies shifts the relative proportion of halos and voids of various sizes. Since the mass function depends on the fully integrated correlation functions (the moments), it is a weaker probe of the three-point function than the bispectrum. However, even current data is sensitive to assumptions about the relevance of higher order moments \cite{Shandera:2013mha, Mantz:2014paa}. Current best fits and 68.3\% confidence intervals from X-ray selected clusters (with {\it Planck} constraints on the homogeneous cosmology and the power spectrum) are $f_{\rm NL}^{\rm loc}=-94^{+148}_{-77}$ for the usual local ansatz and $f_{\rm NL}^{\rm loc}=-48^{+60}_{-11}$ for a scenario where the higher moments are relatively more important than in the standard ansatz \cite{Mantz:2014paa}. Much of the uncertainty in these numbers comes from our inability to fully predict the mass function for visible galaxies and structures from the initial perturbations. Additional closely synchronized simulation and theory work is necessary to remove these additional layers of theory uncertainy. In addition, the constraints are likely to improve considerably as clusters detected via the Sunyaev-Zeldovich effect at higher redshift are added to the sample \cite{Williamson:2011jz, Benson:2011uta, Ade:2013skr, Bleem:2014lea}, and as abundance constraints are considered jointly with clustering constraints \cite{Mana:2013qba}.

\subsection{Uncorrelated Primordial non-Gaussianity, Intermittency and LSS}

 The transition from the coherent inflaton-dominated  state at the end of inflation to an incoherent energy density mix of nonlinear modes gives rise to possibly observable features accessible to CMB and LSS probes. This preheating is a prequel to  the slow relaxation to a fully thermalized particle plasma and the standard model in ways that are far from understood. The scale of the horizon is quite tiny at that the end of inflation, of order a comoving centimetre or so. For there to be observable large scale structure effects, what happens in the preheating transition has to couple to a long wavelength spatial modulation associated with, e.g.,  a coupling constant  \cite{kofman03,zaldarriaga03} or a non-inflaton light (isocon) field \cite{bfhk}.  The generic term for this is modulated preheating (Section~\ref{sec:modreh}), and the curvature fluctuations arising in response are of the local form, $\zeta (x)= F_{NL}(\chi_i (x), g_i (x))$, in  terms of the local initial values just prior to preheating in the tiny horizon volume at that time. Typically the isocon field $\chi_i(x)$ or the coupling ``constant" field $g_i(x)$ would be nearly Gaussian random fields, the latter with a non-zero mean. The simplest possibility is to do an expansion in small values, which leads to linear and quadratic contributions, though statistically independent from the conventional inflaton-induced nearly Gaussian curvature fluctuations. CMB constraints on the size of the quadratic piece ($f^{\rm loc}_{\rm NL,eff}$) are considerably relaxed over the correlated case with its very tight $\fnlloc$ constraints. 
 
Of great interest is where the expansion is not adequate for $F_{NL}(\chi_i (x), g_i (x))$. In a preheating model considered by \cite{bfhk}, $F_{NL}(\chi_i (x))$ was characterized by regularly-spaced positive spikes (looking like an atomic line spectrum), leading to novel structure in spite of its being a local non-Gaussianity. This form held on scales down to the preheating horizon scale, tiny relative to LSS. This was handled in \cite{bfhk} by marginalizing over high spatial frequency modes (about 50 e-foldings worth), resulting in an effective local field map, $F_{\rm NL,eff}(\chi_i (x,R_b))$, with $R_b$ the (LSS) smoothing scale. The form looks like a strongly blended spectrum of lines, i.e.,  one seen at low resolution. Nonetheless the form of such a blended $F_{\rm NL,eff}$ still allows for a wide range of behaviour. A nice way to connect to one's LSS intuition is that $F_{\rm NL,eff}$ is sort of like a nonlinear fuzzy threshold function acting on the underlying Gaussian random field, as occurs in how rare massive clusters are identified as emergent from the initially Gaussian random density field \cite{bbks, bm1}. This leads to biasing, indeed it can be extreme biasing. So it may not surprise that patches of structure can be enhanced. 
 
 A major difference with conventional LSS extreme biasing though is that the field operated on is nearly scale invariant, meaning that very large scales are highly enhanced in curvature. The result is large-scale spatial intermittency, leaving the halo clustering as usual in most places in the Universe, but once in a while a constructive interference would occur between the inflaton-induced fluctuations and the intermittent modulated-preheating-induced fluctuations; i.e., halo clustering would be enhanced over large regions sporadically. The challenge would then be how do you search for such  large scale enhancements. These anomalous events would be rare, so you need a large survey, but you would be trying to identify spatial splotches, suggesting that a power spectrum approach would miss the essence of the effect: a few super-duper-clusters among the ordinary super-clusters. 
 
 Fortunately the large-volume surveys being considered for searching for conventional perturbative non-Gaussianity are also well suited to look for rare intermittent enhancements. It is just that the toolbox used for the search would be different, more local. Another aspect of the large scale intermittency is that a lower resolution survey such as those probing redshifted 21 cm radiation (e.g., CHIME) could be of great utility.  
  
This intermittency is more generic than the specific models computed by \cite{bfhk}.  Near the end of inflation, there is a largely ballistic phase describing the evolution of the fields present, with the familiar stochastic kicks that give rise to the spatial variation of the inflaton highly subdominant over the general drift downward on the potential surface rather than just subdominant. This ballistic phase continues for a time, until nonlinear mode coupling onsets. Once it does, non-equilibrium entropy can be generated in a burst, marked out in time by a randomizing timelike hypersurface, a shock-in-time \cite{bb15}, with a mediation width making it fuzzy. The details vary from model to model.  The shock-in-time framework holds when parametric resonance occurs at the end of inflation, and this is reasonably generic if the longitudinal inflaton potential opens up at the bottom to other transverse degrees of freedom. (For modulation to occur, the transverse walls during inflation should not be so steep as to preclude the long wavelength stochastic fluctuations from having damped due to a high effective mass, i.e., the modulating field must be light.) The shock-in-time picture is not  a useful descriptor if the entropy generation is slow (perturbative preheating), and is modified somewhat if new relatively long-lived energy structures arise in the non-linear phase of preheating which burst forth into entropy only after a delay.  

The growth of curvature in the ballistic phase is tracked by considering how trajectories of nearby points separate from each other or converge towards each other as a function of the difference $\delta \chi_i$ in their initial $\chi_i's$:  $\delta \zeta = [d \ln a/ d \chi ] \delta \chi$ \cite{bfhk,bb15,b2fh}. The trajectory bundles can evolve in a complex form (chaotic billiards was used to describe this process in \cite{bfhk}).The divergence/convergence of the trajectories freezes when the shock-in-time is reached. Of course to really model the process full nonlinear lattice simulations are needed, but this heuristic description gives the essence of the phenomenon \cite{b2fh}. Spikes are associated with trajectory caustics. The modulating coupling constant story is much the same, with trajectory deviation described by $\delta \zeta = [d \ln a/ d g] \delta g$, leading again to a complex highly featured $F_{NL}( g_i (x))$ \cite{bb15}. The common requirement is rapid divergence/convergence of trajectories (Lyapunov growth) and a shock-in-time. 

To relate this type of intermittent non-gaussianity to LSS surveys, accurate mocks of the surveys are needed. One wants to compare the structure generated in models with purely inflaton-induced initial curvature to those with a subdominant preheating-modulation-induced initial curvature in addition. Very large cluster and galaxy catalogues are being constructed using an accelerated version \cite{abbhs15} of the peak patch \cite{bm1} simulation method which accurately reproduces the halo mass function and 2-point halo clustering \cite{abbhs15}. What is contrasted in \cite{abhfs15} is the nonlinear response to 5 initial condition setups: (1) standard Gaussian tilted LCDM (GLCDM), (2) GLCDM with a correlated quadratic non-Gaussianity characterized by  $\fnlloc$, (3)  GLCDM plus an uncorrelated non-Gaussianity characterized by  $f^{\rm local}_{\rm NL,eff}$ (with much larger values allowed by the data), (4) GLCDM plus a single (Gaussian-shaped) $\zeta$-spike in $\chi$ of specified amplitude and width superposed, and (5) GLCDM plus the $F_{\rm NL,eff}(\chi_i (x,R_b))$ arising from a full preheating lattice simulation with many spikes marginalized over high spatial frequencies. These simulations illustrate the range of behaviours described above and, when tailored to be mocks for the specific future LSS experiment under consideration, can be used to optimize the search for the varieties of non-Gaussianity encountered. In particular, for the intermittent varieties (4,5), the focus should be more on searching for rare large-scale events rather than relying on power spectrum or bispectrum measurements. The examples show quite large scale overdensities of clusters and groups can result, a sort of enhanced superclustering over that of GLCDM. 

Very large volume surveys will be automatically suited for the search for such subdominant intermittency, since all one needs is a catalogue with redshift space positions of galaxies or clusters, or low resolution 3D maps such as in CHIME-like intensity mapping experiments. The simulations do show that the extreme bias acting on a nearly scale invariant spectrum is radically different in appearance from the conventional bias acting on the density field, preferentially making large scale splotches that are uncorrelated and act to either add to or subtract from the GLCDM fluctuations around the splotch locations by this random constructive or destructive interference. 

If intermittent anomalies show up in LSS, they should also show up in the CMB, and there has been much discussion about the origin of the large scale anomalies that have been found in the WMAP and Planck data. Another example of rare non-Gaussian intermittency in LSS that would be worthwhile to search for in very large volume surveys is colliding bubbles, if one (or more) happens to have occurred within our accessible Hubble volume. There have been searches for such remnant structures in the CMB.

%%%%%%%%%%%%%%%%%%%%%%%%%%%%%%%%%%%%%%%%%%%%%%%

\section{Potential Systematic Effects}\label{sec:sys}
%Dragan: Astrometric systematics
%Chris: Instrumental
%Donghui: Spectroscopic survey systematics

In order to extract truly primordial non-Gaussianity signature, we must 
understand the observable, the galaxy density contrast to the accuracy that is
required by the galaxy surveys. In this section, we discuss some of the 
theoretical and observational systematics that we must control well enough 
to detect primordial non-Gaussianity from forthcoming galaxy surveys. 

\subsection{Theoretical Systematics}

Although it is true that the galaxies are seeded by the curvature perturbation 
generated in the early Universe, the correlation functions measured from the 
observed galaxy density contrast are wildly different from the primordial one.
For the temperature fluctuations and polarizations of CMB, which are also 
seeded by the same initial perturbations, the fluctuations still 
remain small so that we can model the difference by applying the linear 
perturbation theory with nearly Gaussian statistics; 
the well studied cosmological linear perturbation theory is the key to the 
success of CMB cosmology. 

In linear, Newtonian theory, the observed galaxy density contrast is given by
\be
\delta_{g}({\bf k})
=
(b_1+f\mu^2)^2 \delta_m({\bf k}),
\label{eq:Newtonian_linear}
\ee
including the linear bias factor $b_1$ \cite{1984ApJ...284L...9K}
and the anisotropies ($\mu$ is the 
angle between the Fourier wave-vector and the line-of-sight direction) from 
the redshift-space distortion \cite{1987MNRAS.227....1K}. While
the linear prescription models the galaxy density contrast reasonably well 
on {\it linear} scales ($k_{\rm H}\ll k\lesssim 0.1~{\rm Mpc}/h$ at 
$z\sim0$, here $k_H=a_0H_0$ is 
the wavenumber corresponding to the comoving horizon at present), 
the linear Eq.~(\ref{eq:Newtonian_linear}) must be modified beyond the both 
end.
Understanding these modifications of the galaxy density field from the 
simple linear prediction in Eq.~(\ref{eq:Newtonian_linear}) is the key for 
galaxy surveys to be as fruitful cosmological probes as their CMB counterpart.
In thie section, we discuss current status of modeling 
two of the main theoretical systematics:
the non-linear effects that affect the galaxy density contrast on smaller 
scales and the general relativistic effect that would change the galaxy
density constant on large scales.

\subsubsection{Non-linear systematics}
There are three non-linear effects that alter the observed galaxy density
contrasts from Eq.~(\ref{eq:Newtonian_linear}): non-linear matter clustering,
non-linear redshift space distortion, and non-linear bias. 
Each of them causes a significant deviation of the observed power spectrum 
and bispectrum of galaxies from the leading order predictions; all effects
must be modeled to the accuracy required by surveys. The rule of thumb for 
the current and forthcoming surveys is to make the theoretical prediction 
good to about a percent accuracy.

For surveys targeting galaxies at high ($z\gtrsim1$) redshift, cosmological
non-linear perturbation theory (PT) \cite{Bernardeau:2002la} is available, when
combining the local bias ansatz and the Finger-of-God prescription, to
model the non-linearities in the galaxy power spectrum to the percent level
accuracy at {\it quasi-linear} scales $k\lesssim 0.2\,h/{\rm Mpc}$ at $z\sim2$,
but increases at higher redshifts 
\cite{Jeong/Komatsu:2006,Jeong/Komatsu:2009,Jeong:2010}.
At lower redshifts, however, the non-linearities are so strong and the 
quasi-linear scale, where aforementioned PT can be applied, is quite limited. 
Various techniques to improve upon the PT are suggested:
Renormalization group approach \citep{mcdonald:2007},
Resumming perturbation \citep{matarrese/peitroni:2007},
Renormalized Perturbation Theory \citep{crocce/scoccimarro:2006,montesano/etal:2010,bernardeau/etal:2008},
Closure theory \citep{taruya/hiramatsu:2008},
Lagrangian perturbation theory \citep{matsubara:2008},
TimeRG theory \citep{pietroni:2008},
Effective field theory approach \citep{baumann/etal:2012,carrasco/etal:2012,carrasco/etal:2014,porto/etal:2014,senatore:2014,senatore/zaldarriaga:2014}
and the most recent development based on kinetic theory
\citep{bartelmann/etal:2014,viermann/etal:2014}, and so on.
Some theories have a full description to calculate the non-linear galaxy
power spectrum (see, e.g. \citet{carlson/etal:2013,matsubara:2011}),
but most of them have calculated the non-linear matter power
spectrum; therefore, non-linearities in the galaxy bias 
\cite{2013PhRvD..88b3515S,Schmidt:2013nsa,Desjacques:2013qx,Assassi:2014fva}
and redshift-space distortion \cite{Scoccimarro:2004tg} must be included.
Also, not many calculations have been done for the galaxy bispectrum, for which
we refer the readers to Section~\ref{sec:bispec}.

\subsubsection{General Relativistic effect}

The observed coordinate of a galaxy in the galaxy surveys is given by its 
angular coordinate (RA, Dec) and redshift $z$. We then chart the galaxy
to some physical coordinate based on the assumed background cosmology as 
$(t,{\bf x}) = (\tau(z),\bar{\chi}(z)\hat{\bf x})$, where 
$\tau(z)$ and $\bar{\chi}(z)$ are, respectively, the 
redshift-to-cosmic-time relation and redshift-to-comoving-radius relation
using background cosmology, and $\hat{\bf x}$ is the unit 3D vector 
pointing toward the galaxy. This process assumes that the photons from the galaxy came to us on a  `straight line' (the background geodesic). 
In reality, however, the photon's path is perturbed due to the metric 
perturbations along its way, and we expect a systematic shift
of the observed density contrast relative to the intrinsic one.

In linear, Newtonian theory, the observed galaxy density contrast is 
given by
\be
\delta_{g}({\bf k})
=
(b_1+f\mu^2)^2 \delta_m({\bf k}),
\label{eq:Newtonian_linear}
\ee
including the linear bias factor $b_1$ \cite{1984ApJ...284L...9K}
and the anisotropies ($\mu$ is the 
angle between the Fourier wave-vector and the line-of-sight direction) from 
the redshift-space distortion \cite{1987MNRAS.227....1K}. While
the linear prescription models the galaxy density contrast reasonably well 
on {\it linear} scales ($k_{\rm H}\ll k\lesssim 0.1~{\rm Mpc}/h$ at 
$z\sim0$, here $k_H=a_0H_0$ is 
the wavenumber corresponding to the comoving horizon at present), 
the linear Eq.~(\ref{eq:Newtonian_linear}) must be modified beyond the both 
end.
Understanding these modifications of the galaxy density field from the 
simple linear prediction in Eq.~(\ref{eq:Newtonian_linear}) is the key for 
galaxy surveys to be as fruitful cosmological probes as their CMB counterpart.
In thie section, we discuss current status of modeling 
two of the main theoretical systematics:
the non-linear effects that affect the galaxy density contrast on smaller 
scales and the general relativistic effect that would change the galaxy
density constant on large scales.

In linear theory, Refs. \cite{yoo/etal:2009,yoo:2010,
bonvin/durrer:2011,challinor/lewis:2011,BaldaufEtal,gaugePk,paperI}
have calculated the observed galaxy density contrast 
with the perturbed light geodesic effects that include the 
Sachs-Wolfe effect, integrated Sachs-Wolfe effect, Shapiro time delay, as well
as weak gravitational lensing.
With Gaussian initial conditions, the calculation reads 
(following the notation of \cite{gaugePk})
\begin{equation}
\delta_{\rm g}^{\rm (obs)}({\bf k})
= 
\left(
b + f({\bf k}\cdot\hat{\bf n})^2 + \frac{\cal A}{[k/(aH)]^2} + i\frac{({\bf k}\cdot\hat{\bf n})}{k/aH}{\cal B}
\right) \delta_{\rm m}({\bf k}),
\label{eq:GRterm}
\end{equation}
where $\delta_{\rm m}$ is the matter density contrast and $\hat{\bf n}$ is the 
line-of-sight directional unit vector.
In a $\Lambda$CDM Universe, the coefficients are given by
\begin{eqnarray}
{\cal A}&=& \frac32 \Omega_{\rm m}
\left[
b_{\rm e} 
\left(1 - \frac{2f}{3\Omega_{\rm m}}\right) + 1 + \frac{2f}{\Omega_{\rm m}}
+
{\cal C} - f - 2 {\cal Q}
\right],
\nonumber
\\
{\cal B}&=& f\left[b_{\rm e} + {\cal C} - 1\right],
\nonumber
\\
{\cal C}&=& \frac32 \Omega_{\rm m} - \frac{2}{\bar{\chi}}\frac{1-{\cal Q}}{aH} - 2 {\cal Q}
\nonumber,
\end{eqnarray}
with the matter density parameter $\Omega_{\rm m}$, $f=d\ln D/d\ln a$ is the logarithmic growth of structure  parameter, and $b_{\rm e} = d\ln(a^3\bar{n}_{\rm g})/d\ln a$ and ${\cal Q} = d\ln {\bar{n}_{\rm g}}/d\ln {\cal M}$  are two new parameters that are in principle observable (${\cal M}$ is the magnification).
With primordial non-Gaussianity, one can simply replace 
the linear bias $b$ in Eq.~(\ref{eq:GRterm})
with the non-Gaussian bias 
$b+\Delta b(k)\simeq b + 3(aH/k)^2\Omega_{\rm m}\delta_c(b-1)f_{\rm NL}a/D(a)$ \cite{BruniEtal} .
To simplify the analysis, in Eq.~\ref{eq:GRterm}, we assume a thin radial binning and ignore 
the relativistic correction given by the line-of-sight integration of the 
gravitational potential and its time derivative.

Note that the ${\cal A}$ term scales with wavenumber exactly in the same as the local--type 
primordial non-Gaussianity. We can therefore define the effective $f_{\rm NL}^{\rm loc}$
due to the general relativistic effect as
\begin{equation}
f_{\rm NL}^{\rm (loc) GR}
= \frac{\cal A}{3\Omega_{\rm m}\delta_{c}(b-1)}\frac{D(a)}{a}=\mathcal{O}(1).
\end{equation}
That is, without the local--type primordial non-Gaussianity, we expect the scale--dependent bias 
with amplitude comparable to $f_{\rm NL}^{\rm (loc)}\sim 1$ solely from the 
relativistic kinematics. In order to claim the detection of primordial non-Gaussianity of order 
unity from future galaxy surveys, one therefore must take the full general relativistic correction into account.

For the relativistic correction for the galaxy bispectrum, a similar 
calculation must be done in the second--order perturbation theory because
the leading order contribution to the galaxy bispectrum comes from correlating
one second order density contrast to two linear $\delta_g$'s. The calculation
of the second order relativistic corrections has been done by 
\cite{2ndorder1,2ndorder2,2ndorder3}, but the effective non-Gaussianity 
is still yet to be computed.

\subsection{Observational Systematics}

Future large-scale structure surveys have a chance at measuring primordial 
non-Gaussianity at an unprecedented precision, $\sigma(\fnlloc)\simeq 1$. 
Sensitivity to such tiny signatures will require correspondingly stringent 
control of systematic errors.

The fundamental observable quantity in studies of LSS clustering is the galaxy density
contrast $\delta_{\rm g}({\bf x}) = N_{\rm g}({\bf x})/\bar{N}_{\rm g}({\bf
  x})-1$, fractional fluctuation in the galaxy number relative to the expected
mean $\bar{N}_{\rm g}({\bf x})$.  Any error in estimating $\bar{N}_{\rm
  g}({\bf x})$ therefore causes a systematic bias in measuring the density
contrast, which consequently propagates into the galaxy power spectrum and bispectrum, and
then to the cosmological parameters like those describing the primordial non-Gaussianity.  

The principal systematic errors afflicting LSS measurements have diverse
origin; some examples are: imperfectly known response of the instrument,
impact of baryons on small-scale clustering, Galactic dust, atmospheric
blurring and extinction, and (for photometric surveys) redshift
errors. Because future measurements of primordial non-Gaussianity will rely
both on large spatial scales (especially in the power spectrum measurements)
and small scales (in the bispectrum measurements), control of systematics will
need to encompass a diverse set of tools and techniques.  For the measurement
of the scale-dependent bias in the galaxy power spectrum, any systematic that
exhibits large-scale variation will partially mimic the primordial non-Gaussianity signal, as long as
that error departs from the exact $k^{n_{\rm s}}$ scaling at large scales. One
particularly well-studied example is the Galactic extinction; the angular
correlation function of the Galactic dust fluctuations scales as
$C_\ell\propto \ell^{-2.5}$ on degree scales at high galactic latitudes
$|b|>45\,^\circ$ \cite{SFD}.  Similar caution is necessary in regards to the
bispectrum; one must first understand the three-point function signatures of the
systematic errors in order to isolate the primordial non-Gaussianity signal.

A particularly general class of systematics are the {\it photometric
  calibration errors} \citep{Vogeley_sys,Scranton_sys,Ross:2011cz,
  Pullen:2012rd,Ho:2012vy, Huterer:2012zs,Agarwal:2013ajb,Leistedt:2013gfa} --
these are the systematic that effectively causes the magnitude limit of the
sample to vary across the sky, typically biasing the true galaxy power
spectrum at {\it large} spatial scales. Some of the common calibration errors
are unaccounted-for extinction by dust or the atmosphere, obscuration of
galaxies by bright stars, and varying sensitivity of the pixels on the camera
along the focal plane.  A rough estimate of the required control on the
calibration error can be obtained as follows: on the largest spatial scales
($k\simeq H_0$), the density fluctuation is $\delta\rho/\rho\simeq 5\times
10^{-5}$. 
%\dragan{This is $(\Delta^2(k=H_0, z=0))^{1/2}$}. 
The primordial non-Gaussianity affects the bias of dark matter halos, which at
these scales, and for $\fnlloc\simeq 1$ and typical bias $b\simeq 3$, is $\Delta
b\simeq 5$. Therefore, a Hubble-volume survey would need to be sensitive to
fluctuations in the number density of objects of order 
%\dragan{(perhaps we quote a slightly larger number 0.001 listed below in this
%  inlined equation?)}
\begin{equation}
\left (\frac{\delta n}{n}\right )_{\rm calib}\simeq 
\Delta b\left (\frac{\delta\rho}{\rho}\right )\simeq 3\times 10^{-4}.
\end{equation}
This very challenging requirement is supported by more detailed calculations,
which indicate that photometric calibration needs to be understood at the
$10^{-3}$ (or $\simeq 0.1\%$) level \cite{Huterer:2012zs}, this result
depends fairly strongly on the faint end of the luminosity function which
effectively converts the calibration fluctuations into variations in the
measured galaxy counts.  While apparently extremely stringent, the required
understanding of the photometric calibration can be addressed and, hopefully,
achieved in the near future with the combination of careful observation and
analysis, as well as partial self-calibration (internal measurement) of the
parameters that describe these systematics
\citep{Ho:2012vy,Agarwal:2013ajb,Shafer:2014wba}.

Good theoretical understanding of galaxy clustering at {\it small} spatial
scales (between 1$\hinvmpc$ and 10$\hinvmpc$) is crucial for measurements of
the LSS bispectrum, which has a nice feature of being sensitive to all shapes
of primordial non-Gaussianity. The number of available modes in a survey goes
as $k_{\rm max}^3$, where $k_{\rm max}$ is the maximum wavenumber probed;
hence a huge amount of additional information is available provided small
scales can be modeled reliably. The principal uncertainties are caused by
the combined baryonic and nonlinear effects that modify the clustering at
small scales. Here, a combination of hydrodynamical simulations and direct
observations of the mass density profiles (e.g.\ via gravitational lensing)
will be key to achieve sufficient understanding of the small-scale
clustering. As in the case of large-scale systematics, self-calibration via
flexible modeling of the remaining systematics
\citep{Mohammed:2014lja,Bielefeld:2014soa}, and projecting out information
that is affected by the systematics while keeping the bulk of the cosmological
information \citep{Leistedt:2013gfa,Leistedt:2014wia,Eifler:2014iva}, can both
be very effective.

Surveys that do not have spectroscopy of the collected objects typically rely
on the photometric redshifts of the source galaxies in order to get
information about the full three-dimensional distribution of structure in the
universe. The {\it photometric redshift errors} typically blur and even
rearrange clustering in the radial direction, and therefore represent a
potentially major concern in LSS surveys seeking to constraint primordial
non-Gaussianity through the clustering of sources. As with the other
systematics, the relevant quantity that impacts the result is the
unaccounted-for error in the photometric redshift distribution (the ``error in
the error''). The typical redshift error should be known at roughly the level
of $\delta z_p\simeq 0.003$, which can be achieved
with multi-band photometry \citep{fnlsurvey}. A particular concern are the large individual
(``catastrophic'') photometric errors, where a galaxy's true redshift is badly
misestimated where again the redshift is biased or absent due to a
number of possible reasons. These catastrophic errors need to be understood at
a level of roughly 0.1\% \citep{Bernstein:2009bq,Hearin:2010jr}.
%\dragan{(number for DE; I am guessing also true for PNG?)}

{\it Redshift failures} represent a separate challenge for spectroscopic
surveys \cite{Cunha:2012us,Newman:2013cac}.  To observe the high-redshift
galaxies, future spectroscopic surveys such as HETDEX and Euclid are targeting
a single bright line, such as Ly$\alpha$ or H$\alpha$; these lines can be
confused by the lower redshift oxygen lines. There fortunately exist known
astrophysical methods, e.g.\ using the equivalent width cut, to clean up the
contamination from low-redshift interlopers. These methods are proven to work
at a percent level, leading to little or no bias on the baryon acoustic
oscillation measurement and the dark energy constraints, the main
science drivers of these surveys.  For the primordial non-Gaussianity measurement, however, more
stringent systematic control is required as the signal mostly comes from the
larger scales, where the systematic effects from power spectrum of interlopers
can be larger than the targeted galaxy power spectrum.

%\subsection{Astrophysical Systematics} 

\section{Conclusion}

%The obvious conclusion of our workshop is that the motivation to probe inflation with large-scale structures are stronger than ever. In
%the aftermath of the success of a generation of CMB experiments that culminated with the revolutionary constraints obtained by WMAP and Planck, large-scale structure the aftermath of Planck's success, these surveys hold the greatest potential to constrain the shape of the primordial spectrum and to single out the physics of inflation via the measurement of non-Gaussianity. We emphasized the existence of clear and meaningful theoretical targets which  are within reach of currently planned surveys. We have no doubt the importance of these targets will  motivate the community effort required to tackle the modeling, observational and analysis challenges we identified.
The obvious conclusion of our workshop is that the motivation to probe inflation with large scale structures are stronger than ever. The success of a generation of CMB experiments has culminated with the revolutionary constraints obtained by WMAP and Planck. Looking forward, LSS surveys hold the greatest potential to improve constrains on the shape of the primordial spectrum and to single out the physics of inflation via the measurement of non-Gaussianity. We emphasized the existence of clear and meaningful theoretical targets which  are within reach of currently planned surveys. We have no doubt the importance of these targets will  motivate the community effort required to tackle the modeling, observational and analysis challenges we identified.

\bibliographystyle{ieeetr}
%\bibliography{biasobs,NGotherProbes,matt_ng_white_paper,multifield2,futuresurvey.bbl}
%\bibliography{futuresurvey}
%\bibliography{BS}

\begingroup\raggedright

\end{document}